\documentclass[%
 reprint,
 amsmath,amssymb,longbibliography,
 aps
]{revtex4-1}

\usepackage{graphicx}
\usepackage{dcolumn}
\usepackage{bm}

\usepackage{braket}

\def\bcen{\begin{center}}
\def\ecen{\end{center}}
\usepackage{epstopdf}
\usepackage{soul}
\usepackage[breaklinks=true,colorlinks]{hyperref}
\begin{document}

\preprint{APS/123-QED}

\title{A review on quantum information processing in cavities}

\author{Nilakantha Meher$^1$}
\email{nilakantha.meher6@gmail.com}
\author{S. Sivakumar$^2$}%
 \email{sivakumar.srinivasan@krea.edu.in}
\affiliation{%
 $^1$Department of Chemical and Biological Physics,
Weizmann Institute of Science, Rehovot 7610001, Israel\\
$^2$Division of Sciences, Krea University, Andhra Pradesh 517646, India
}%




\date{\today}
\begin{abstract}
The processing of information and computation is undergoing a paradigmatic shift since the realization of the enormous potential of quantum features to perform these tasks. Coupled cavity array is one of the well-studied systems to carry out these tasks. It is a versatile platform for quantum networks for distributed information processing and communication. Cavities have the salient feature of retaining photons for longer durations, thereby enabling them to travel coherently through the array without losing them in dissipation. Many quantum information protocols have been implemented in arrays of coupled cavities. These advancements promise the suitability of cavity arrays for large scale quantum communications and computations. This article reviews a few theoretical proposals and experimental realizations of quantum information tasks in cavities.
\end{abstract}

\pacs{Valid PACS appear here}
\maketitle


\section{Introduction}\label{sec1}
Quantum theory was formulated in response to the failure of classical physics to explain observed phenomena such as the black-body radiation, specific heat at low temperatures, diffraction pattern due to cathode rays, etc. The domain of its applicability expanded from atoms and molecules to matter in bulk, and beyond including cosmic scale objects. Within a short span of time, physicists realized that the framework raises very deep questions about our conception of nature. Some of
the fundamental issues are related to the idea of locality \cite{Einstein1935PR,Bell1964PPF}. It requires fine experiments to settle the issues arising out of the framework. Testing of Bell's inequality using polarization entangled
photons is one of the examples of such attempts \cite{Aspect1982PRL}. An achievement of such fundamental studies is the emergence of quantum information as a discipline
\cite{DiVincenzo2000FP,Feynman1982IJTP,Benioff1980JStatPhys,Ladd2010Nature}. 
This is a generic term to indicate processes that make explicit use of quantum principles to overcome the limitations arising out of the classical framework. Protocols for teleportation \cite{Bennet1993PRL}, quantum state transfer \cite{Cirac1997PRL,Cirac1998PhysScr}, dense coding \cite{Bennett1992PRL}, quantum cryptography \cite{Bennett1992JourCrypt} are some of the fine achievements of these studies. A definite requirement for the realization of such protocols is ability to manipulate and transfer quantum bits (qubit, in short). A qubit is an arbitrary superposition of two classical bits $\ket{0}$ and $\ket{1}$ that can carry information \cite{Schumacher1995PRA,Neil,Pathak}. Two orthogonal states of a quantum system, for example, two chosen energy levels of an atom \cite{Saffman2010RevModPhys,Saffman2016JPhyB}, polarization states of photons \cite{Torma1996PRA}, spin states of electrons \cite{Loss1998PRA}, etc. are being used as qubits for realizing quantum information protocols. Among them, photons are the excellent carrier of quantum information as they do not interact with each other and propagate over large distances \cite{Delbecq2013NatComm}.  They are being used as flying qubits for quantum communication in free space
 \cite{Bouwmeester1997Nature,Naik2000PRL,Tittel2000PRL,Duan2011Nature,
Liao2017NatPhotonics}.

Recently, realizing quantum information protocols using photons as the information carrier in coupled cavity arrays has received a lot of attention \cite{Northup2014NatPhy,Perez2013PRA,Perez2013PRAFeb, Chapman2016NatComm,Bose2007JModOpt,Giampaolo2009PRA,Angelakis2007JOSAB,LiJian2008CommThPhys,
Almeida2016PRA,LiJian2010CommThPhy,Liew2012PRA,Liew2013NJP,Lin2009APL,
YangLiu2015NJP,Meher2017ScRep,Li2019OptExp,Mendonca2020PRA}. This is because a coupled cavity array allows a high degree of controllability and scalability \cite{Notomi2008NatPhy}. Importantly, each cavity of the array is addressable and the state of a selected cavity can be manipulated. A cavity consists of two highly reflecting mirrors that trap the photons for a fraction of a second \cite{VahalaBook,HarocheBook,Tanabe2007NatPhotonics,
Kavokin2017Book,Hood2001PRA}. The lifetime of photons inside the cavity is proportional to the quality factor ($Q$ factor) of the cavity \cite{Fox}, that is, $\tau\sim\frac{Q}{\omega}$, where $\omega$ is the resonance frequency of the cavity. A cavity with large $Q$ factor stores photons for a long time. With the current technology, high values of $Q$ are achievable in photonic crystal cavities \cite{Vuckovic2001PRE,Reese2001JourVacSc,Srinivasan2002OptExpress,
Srinivasan2003APL,Akahane2003Nature,Takahashi2007OptExp,Majumdar2012PRA}, Fabry-Perot cavities \cite{Rempe1992OptLett,Hood2001PRA,Hunger2010NJP}, superconducting resonators \cite{Blais2020NatPhys,Blais2004PRA}, toroidal microdisks \cite{Armani2003Nature,Beyoucef2008PRB}, whispering gallery modes of silica and quartz microspheres \cite{Gorodetsky1996OptLett,Valarie1997MatScEngB,Vernooy1998PRA,Buck2003PRA}, etc. Reported experimental values of $Q$ in such cavities are listed in Table. \ref{ExpParameter}. The details on fabrication of various type of cavities can be found in the Refs. \cite{VahalaBook,Vahala2003Nature,Noda2016OpFCConf}.

The field in a cavity couples to the field in a nearby cavity, for example, due to evanescent fields \cite{Yariv1999OL,Hartmann2016JOpt}. The coupling strength between the cavities, denoted by $J$, depends on the distance between the cavities through $J \propto e^{-d/k}$ \cite{Matthieu2012OptLett,Perez2013PRA,Siegle2017LSA}, where $d$ is the distance of separation between the cavities and $k$ is a constant.  Because of the coupling, the cavities exchange photons with each other. A series of cavities arranged in close proximity so as to exchange photons among them forms a cavity array. In fact, Notomi \textit{et al.} \cite{Notomi2008NatPhy} reported an array of about 100 cavities each of the size of the wavelength of the field. High-$Q$ cavities are essential not only for fundamental studies \cite{Haroche1991PhysicsWorld,Kofman1996PRA,Kurizki2015PNAS, Hwang2016PRL,Greentree2006NatPhy,
Hartmann2006NatPhy,Meher2016JOSAB,Meher2016Conf,Liu2015IJTP} but also for many applications such as imaging of atoms beyond diffraction limit \cite{Mabuchi1996OptLett,Rempe1995AppPhyB}, atom-cavity microscope \cite{Hood2000Science}, controlling light pulse propagation \cite{Shimizu2002PRL}, bio-sensor \cite{Vollmer2002APL, Cooper2002NatRevDrugDisc}, optical sensor \cite{Bitarafanm2017Sensors,Qiao2018MicroMachines,Krioukov2002OptLett}, quantum heat engine \cite{Bergenfeldt2014PRL,Niedenzu2018NatComm, Hardal2015ScRep,Gelbwaser2013EPL,Gelbwaser2013PRA,
Dodonov2018JPhysA,Ghosh2017PNAS,Jiteng2021ScAdv}, photonic transistor \cite{Hu2017ScRep,Chen2013Science}, quantum memory \cite{Maitre1997PRL,Giannelli2018NJP}, quantum cloning machine \cite{Milman2003PRA,Zou2005PRA}, entanglement detection \cite{Zhou2014PRA,Zhou2015PRA,Sheng2013CSB}, etc.  

In recent years, extensive studies have looked at employing cavities to investigate fundamentals of physics as well as to implement some of the quantum mechanical applications. Many exciting features of coupled cavity arrays have been explored theoretically as well as experimentally. In this review, we mostly focus on the implementation of various quantum information protocols in cavity arrays \cite{Bose2007JModOpt,Giampaolo2009PRA,Angelakis2007JOSAB,LiJian2008CommThPhys,
Almeida2016PRA,LiJian2010CommThPhy,Liew2012PRA,Liew2013NJP,Lin2009APL,
YangLiu2015NJP,Meher2017ScRep,Li2019OptExp,Mendonca2020PRA}. In particular, we discuss the possibility of quantum state transfer through the array, generation of various entangled states between the cavities, demonstration of quantum gates, realization of quantum teleportation and quantum dense coding between distant cavities. Quantum state transfer, teleportation and dense coding are essential to realize quantum communication, whereas quantum gates are the basic elements for quantum computation. 
\begin{table}[t]
\begin{center}
\caption{Reported experimental values of $Q$ factor of cavities in various materials.}
 \begin{tabular}{| c |c |  c |} 
 \hline
 System  & $Q$ factor & References \\ [0.5ex] 
 \hline\hline
 Photonic crystal cavity  & $10^5-10^6$ & \cite{Tanabe2007NatPhotonics,Takahashi2007OptExp,Notomi2008NatPhy,
 MajumdarRund2012PRB}\\ [1ex] 
  \hline
 Toroidal microresonator   & $ 10^8$ & \cite{Armani2003Nature}\\ [1ex]
 \hline
 Superconducting resonator  & $10^4-10^7$ & \cite{Brune1996PRL, Frunzio2005IEEE}\\[1ex]
 \hline
  Fused-silica microspheres  & $ 10^{10}$ & \cite{Gorodetsky1996OptLett,Vernooy1998PRA}\\[1ex]
 \hline
\end{tabular}
\label{ExpParameter}
\end{center}
\end{table}

\section{Single cavity}
\subsection{Quantization of electromagnetic field inside a cavity}
Many experiments in optics are explainable by treating the electromagnetic field classically \cite{Meystre2007Spr}. However, quantization of the electromagnetic field is essential to explain a few notable experimental outcomes, such as the black-body spectrum, spontaneous emission, Lamb shift, resonance fluorescence, squeezed light, etc \cite{Scully1997Book}. Here, we briefly discuss the quantization of electromagnetic field in a cavity.
\begin{figure*}
\begin{center}
\includegraphics[height=6cm,width=8cm]{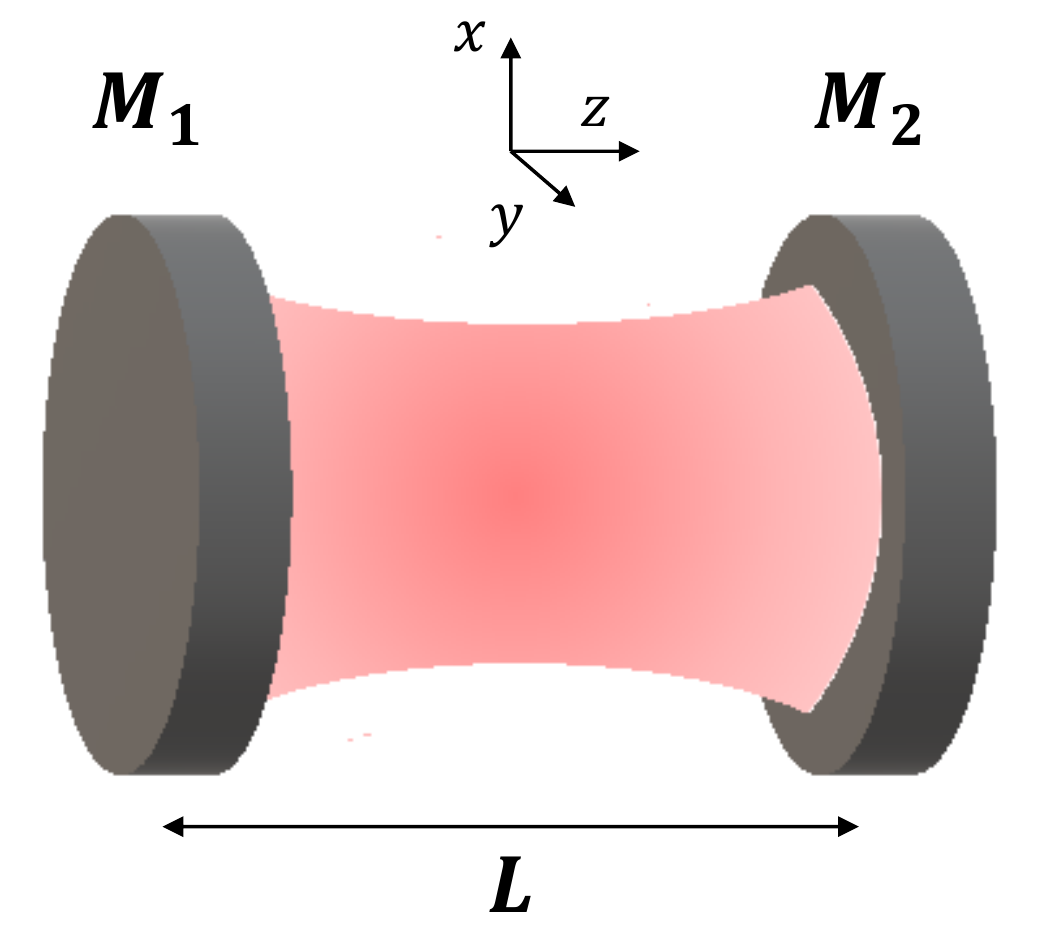}
\caption{A schematic of a cavity which is a pair of two mirrors $M_1$ and $M_2$ separated by a distance $L$. Field is propagating along $z$-direction and electric field is oscillating along $x$-direction. }
\label{EmptyCavity}
\end{center}
\end{figure*}

The space-time evolution of the electromagnetic field in vacuum is described by the Maxwell's equations, which are
\begin{subequations}\label{Ch1maxwelleqn}
\begin{eqnarray}
&\vec{\nabla} \cdot \vec{E}(\vec{r},t)=0,\\
&\vec{\nabla} \cdot \vec{B}(\vec{r},t)=0,\\
&\vec{\nabla} \times \vec{E}(\vec{r},t)=-\frac{\partial}{\partial t}\vec{B}(\vec{r},t),\\
\text{and}~~~~~~~ &\vec{\nabla} \times \vec{B}(\vec{r},t)=\mu_0 \epsilon_0 \frac{\partial}{\partial t} \vec{E}(\vec{r},t),
\end{eqnarray}
\end{subequations}
where $\mu_0$ and $\epsilon_0$ are respectively the permeability and permittivity of the free space. A consequence of these equations is that the electric and magnetic fields satisfy  
\begin{align}\label{Ch1waveeqn}
\vec{\nabla}^2\vec{X}(\vec{r},t)=\frac{1}{c^2}\frac{\partial^2}{\partial t^2} \vec{X}(\vec{r},t),
\end{align}
so called wave equation, where $\vec{X}(\vec{r},t)$ is $\vec{E}(\vec{r},t)$ or $\vec{B}(\vec{r},t)$. These equations imply that the electromagnetic wave propagates in free space with speed $c$. Another consequence of Eqns. (\ref{Ch1maxwelleqn}$a$)-(\ref{Ch1maxwelleqn}$d$) is that the fields $\vec{E}$ and $\vec{B}$ are transverse, \textit{i.e.}, $\vec{E}$, $\vec{B}$ and the direction  of propagation are mutually  perpendicular to each other. These equations can be modified to include charges and currents as well. If boundary conditions are imposed on the fields, these equations can describe the fields in a confined geometry.

Consider the electromagnetic field between two perfectly conducting plates separated by a length $L$  (cavity) as shown in Fig. \ref{EmptyCavity}. In the limit of large $L$, the field corresponds to the electromagnetic field in free space. The field is assumed to be propagating along $z$-direction and  the electric field is polarized in $x$-direction, \textit{i.e.}, $\vec{E}(r,t)=\hat{e}_xE_x(z,t)$, where $\hat{e}_x$ is the polarization direction. As the walls are perfectly conducting, the electric field vanishes at the boundaries at $z=0$ and $z=L$. In order to write the explicit form of $\vec{E}$ for the cavity field, consider the fundamental modes of the electromagnetic field, which are the eigenfunctions of the spatial part of the wave equation.  Any arbitrary distribution of the electric field inside the cavity can be expressed as a linear combination of these fundamental modes \cite{Scully1997Book},
\begin{equation}\label{electric}
E_x(z,t)=\sum_{j=1}^\infty A_j q_j(t)\sin(k_j z),
\end{equation}
where $q_j$ is the amplitude of the $j$th fundamental mode with the dimension of length and $k_j=j\pi/{L}$ is the magnitude of the wave vector. The amplitude $q_j$ plays the role of the canonical position for an oscillator. The expansion coefficient $A_j=\left(\frac{2\omega_j}{V\epsilon_0}\right)^{1/2}$ where $\omega_j$ is the frequency of $j$th fundamental mode and $V$ is the modal volume. These modes satisfy the orthogonality relation
\begin{align}
\int_{0}^L \sin (k_n z) \sin (k_m z) dz=\frac{L}{2}\delta_{nm}.
\end{align} 
Boundary conditions on the electric field restrict the possible frequencies to
\begin{align}\label{Ch1allowedfrequency}
\omega_j=\frac{j\pi c}{L}.
\end{align}
These are the resonance frequencies of the cavity. The separation between two successive resonance frequencies is $\pi c/L$, which is negligible if $L$ is large. Similarly, the magnetic field inside the cavity is
\begin{equation}\label{magnetic}
B_y(z,t)=\sum_j A_j\left(\frac{p_j(t)\epsilon_0\mu_0}{k_j}\right)\cos(k_j z).
\end{equation}
Here $p_j(t)=\dot{q}_j(t)$ is analogous to the canonical momentum for a particle in the Hamiltonian dynamics.
The Hamiltonian for the electromagnetic field is
\begin{equation}
H=\frac{1}{2}\int dV\left[\epsilon_0 E_x^2(z,t)+\frac{1}{\mu_0}B_y^2(z,t)\right].
\end{equation}
Using the expressions for $E_x$ and $B_y$ respectively from Eqns. (\ref{electric}) and (\ref{magnetic}), the total Hamiltonian becomes
\begin{equation}
H=\frac{1}{2}\sum_j\left(p_j^2(t)+\omega_j^2~ q_j^2(t)\right),
\end{equation}
which has the same structure as that for a set of independent harmonic oscillators. In essence, each fundamental mode of the electromagnetic field is equivalent to an oscillator. The electric field and the magnetic field are equivalent to the position and the momentum of the oscillator respectively.

Now, the quantization of the electromagnetic field becomes straightforward.  Quantization provides an elegant way of understanding the classical wave picture of the field in terms of quantum picture.  It can be inferred from the wave equation given in Eqn. (\ref{Ch1waveeqn}) that the respective amplitudes, namely, $q$ and $p$ of the electric and magnetic fields obey the classical equations of motion of a harmonic oscillator. The canonical variables $q(t)$ and $p(t)$ are represented by self-adjoint operators $\hat{q}$ and $\hat{p}$ which satisfy the commutation relation $[\hat{q},\hat{p}]=i\hbar \hat{I}$. For further analysis, it is advantageous to define
\begin{align}\label{Ch1creationannihilation}
\hat{a}_j=\frac{1}{\sqrt{2\hbar\omega_j}}(\omega_j \hat{q}_j+i\hat{p}_j),\nonumber\\
\hat{a}_j^\dagger=\frac{1}{\sqrt{2\hbar\omega_j}}(\omega_j \hat{q}_j-i\hat{p}_j),
\end{align}
which satisfy $[\hat{a}_j, \hat{a}_k^\dagger]=\hat{I}\delta_{j,k}$. The operators $\hat{a}_j$ and $\hat{a}_j^\dagger$ are called the creation and annihilation operators respectively and $\hat{I}$ is the identity operator. In terms of these operators, the Hamiltonian for the quantized electromagnetic field is
\begin{equation}
\hat{H}=\sum_j \hbar\omega_j\left(\hat{a}_j^\dagger \hat{a}_j+\frac{1}{2}\right).
\end{equation}
The term $\hbar\omega_j/2$ corresponds to the energy of the vacuum field of the $j$th mode. The term $\hat{a}_j^\dagger \hat{a}_j$ represents the number operator for $j$th mode.
\subsection{Single-mode field in a cavity}
As noted in the previous subsection, each mode of the electromagnetic field inside a cavity is equivalent to a harmonic oscillator. These independent modes are described in their respective Hilbert spaces.  A single-mode electromagnetic field has a specific spatial distribution of the electric field decided by the geometry of the cavity and the boundary conditions. The Hamiltonian for a single-mode field is
\begin{equation}
\hat{H}= \hbar\omega\left(\hat{a}^\dagger \hat{a}+\frac{1}{2}\right).
\end{equation}
The eigenvectors of this Hamiltonian are $\{\ket{n}\}$ with corresponding energy eigenvalues $\{E_n=(n+1/2)\hbar\omega\}$. These eigenstates are called the number states or Fock states, the most fundamental quantum states of a single-mode electromagnetic field. The state $\ket{n}$ represents the electromagnetic field having $n$ photons. A consequence of the quantization of electromagnetic field is that the vacuum state has non-zero energy, a feature which has no classical counterpart. The state $\ket{0}$ represents the vacuum state with energy $E_{0}=1/2\hbar \omega$. Being the eigenstates of the self-adjoint operator $\hat{H}$, the number states $\{\ket{n}\}$ form a complete basis for the Hilbert space associated with the single-mode field. Hence, any state of that single-mode field can be expressed as a superposition of the number states.  Suitable superpositions of these number states generate many important quantum states such as the coherent states \cite{Glauber1963PR,Glauber1963PR2}, Schr\"{o}dinger-cat states \cite{Dodonov1974Physics}, squeezed states \cite{Slusher1985PRL,Wu1986PRL}, photon-added coherent states \cite{Agarwal1991PRA}, displaced number states \cite{Oliveira1990PRA}, number state filtered coherent states \cite{Meher2018QInP,Meher2020QInP}, nonlinear coherent states \cite{Sivakumar1999JPhysA,SIVAKUMAR1998PLA,Sivakumar2000JOptB,
Sivakumar2000JPhysA}, etc. Preparation of these nonclassical states in cavities is of current interest because of their usefulness in basic studies, quantum information applications \cite{Cochrane1999PRA,Jeong2002PRA, Oliveira2000PRA} and quantum metrology \cite{Caves1981PRD,Birrittella2012PRA,Meher2020QInP,Tan2014PRA}. Many theoretical proposals have been suggested to generate Fock states with arbitrary number of photons \cite{Filipowicz1986JOSAB,Krause1987PRA,
Krause1989PRA,Meystre1987OptLett,Cummnings1989PRA,
Slosser1989PRL,Harel1996PRA,Weidinger1999PRL,
Kuhn1999AppPhyB,Franca2001PRL,Keller2004Nature,Cosacchi2020PRR,
Krastanov2015PRA,Brown2003PRA,Uria2020PRL}. Fock states with photon number up to $n=1$ \cite{McKeever2004Science}, 2 \cite{Varcoe2000Nature,Brattke2001PRL,Bertet2002PRL}, 7 \cite{Zhou2012PRL} and 15 \cite{Wang2008prl} have been reported to have been realized in cavity experiments. Some of the theoretical proposals include generation of squeezed states \cite{Meystre1982PLA,Monteiro2005PhyA,Lutterbach2000PRA,Werlang2008PRA}, superposition of coherent states \cite{Domokos1994PRA,Szabo1996PRA,VillasBoas2003PRA,
Zeng1998PhyLettA,Plastina1999EPJD}, arbitrary  superposition of number states \cite{Vogel1993PRL,Garraway1994PRA,Kozhekin1995PRA,
Parkins1993PRL,Parkins1995PRA,Law1996PRL,Zheng2006PRA,Rojan2014PRA}, number state filtered coherent state \cite{Meher2018QInP}, etc. in cavities. However, experimental generation of few of these states such as squeezed states \cite{Ourjoumtsev2011Nature} and superposition of coherent states \cite{Brune1992PRA,Brune1996PRL,Vlastakis2013Science} in cavities have been reported. An experimental challenge is to retain the states for long time, made difficult due to the finite lifetime of photons inside the cavity and hence, a feedback mechanism is necessary to retain the nonclassical states for a long time \cite{Varcoe2000Nature,Sayrin2011Nature,Vitali1998PRA}.

\subsection{Cavity Quantum Electrodynamics (QED)}
A single atom in free space interacts with a continuum of modes of electromagnetic field. If the atomic and field frequencies match, then there is a high probability of photon absorption by the atom.  On de-excitation from a higher energy level to a lower energy level, the atom emits a photon. This is an irreversible process in free space. The rate of emission is decided by the density of modes of the field \cite{PurcelEMl1946PR}. In three dimensions, the density of modes is proportional to $\omega^2$, where $\omega$ is the frequency of electromagnetic field mode \cite{Fox}. Interaction between the atom and the field can be tailored by modifying the mode density, which is possible by placing the atom in a cavity. The direction and the rate of spontaneous emission from an atom in a cavity can be controlled \cite{PurcelEMl1946PR,Solomon2000phystatsolidi,Gerard1998PRL}, and is changeable by tuning the cavity resonance frequency and the atom-field coupling strength \cite{Englund2005PRL,
Goy1983PRL,Hulet1985PRL,Gabrielse1985PRL,Heinzen1987PRL,
Gerard1998PRL,Bayer2001PRL,Solomon2000phystatsolidi,Lodahl2004Nat}. These properties have been used for designing nanocavity laser \cite{Altug2006NatPhys,Altug2005OptExp}, quantum encryption \cite{Matsueda1995LasEleOpt}, etc. 

\begin{figure}
\begin{center}
\includegraphics[height=4.5cm,width=7cm]{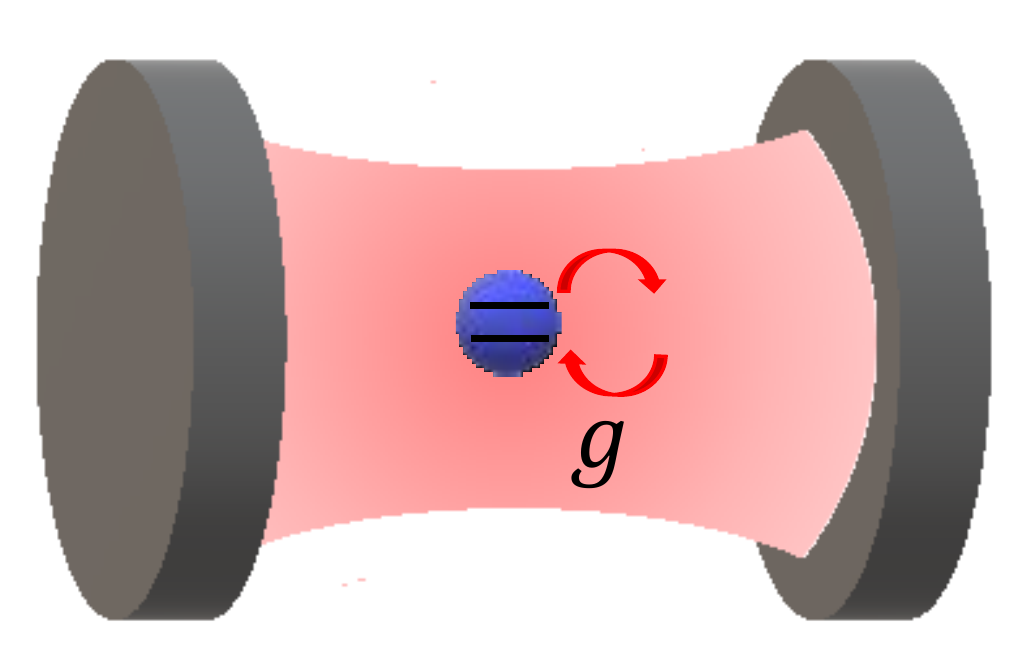}
\caption{A schematic of a cavity coupled to a two-level atom with the coupling strength $g$.}
\label{AtomCavity}
\end{center}
\end{figure}

\begin{figure}
\begin{center}
\includegraphics[height=6.2cm,width=8cm]{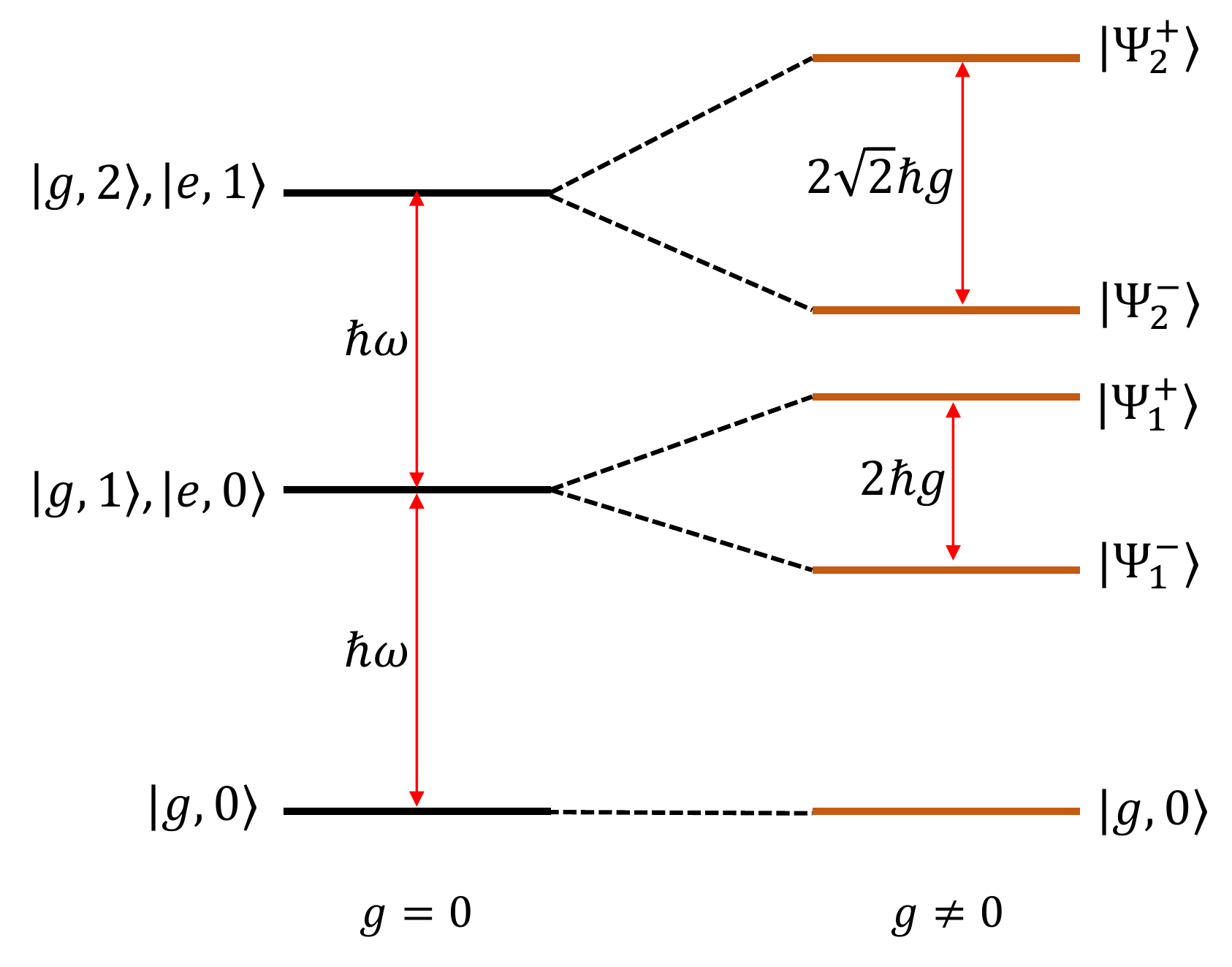}
\caption{Few lowest energy levels of atom-cavity system for $g=0$ (left) and $g\neq 0$ (right) with zero detuning.}
\label{EnergylevelsJC}
\end{center}
\end{figure}

\begin{table*}
\begin{center}
\caption{Reported experimental parameters in atom-cavity system}
 \begin{tabular}{|| c | m{2.5cm} | m{2.5cm} | m{2.5cm} ||} 
 \hline
 References & Atom-cavity coupling strength $(g/2\pi)$ & Atomic decay rate $(\gamma/2\pi)$ & Cavity decay rate $(\kappa/2\pi)$ \\ [0.5ex] 
 \hline\hline
 Turchette \textit{et.al}\cite{Turchette1995PRL} & 20 MHz & 2.5 MHz & 75 MHz \\
 McKeever \textit{et.al}\cite{McKeever2003Nature} & 16 MHz & 2.6 MHz & 4.2 MHz \\
 Birnbaum \textit{et.al} \cite{Birnbaum2005Nature} & 34 MHz & 2.6 MHz & 4.1 MHz \\
 Hennrich \textit{et al.}\cite{Hennrich2005PRL} & 2.5 MHz & 3 MHz & 1.25 MHz \\
 Boozer \textit{et.al}\cite{Boozer2006PRL} & 34 MHz & 2.6 MHz & 4.1 MHz \\
Aoki \textit{et al.}\cite{Aoki2006Nature} & 50 MHz & 2.6 MHz & 17.9 MHz \\
Hijlkema \textit{et.al}\cite{Hijlkema2007NatPhys} & 5 MHz & 3 MHz & 5 MHz \\
Fortier \textit{et.al}\cite{Fortier2007PRL} & 17 MHz & 6 MHz & 7 MHz \\
Dayan \textit{et.al}\cite{Barak2008Science} & 70 MHz & 2.6 MHz & 165 MHz \\
Terraciano \textit{et.al}\cite{Terraciano2009NatPhys} & 1.5 MHz & 6 MHz & 3.2 MHz \\
Mucke \textit{et.al}\cite{Muecke2010Nature} & 4.5 MHz & 3 MHz & 2.9 MHz \\
Specht \textit{et.al}\cite{Specht2011Nature} & 5 MHz & 3 MHz & 2.5 MHz \\
Koch \textit{et al.}\cite{Koch2011PRL} & 16 MHz & 3 MHz & 1.5 MHz \\
Zhang \textit{et al.}\cite{Zhang2011PRA} & 23.9 MHz & 2.6 MHz & 2.6 MHz \\
Ritter \textit{et.al}\cite{Ritter2012Nature} & 5 MHz & 3 MHz & 3 MHz \\
Reiserer \textit{et.al} \cite{Andreas2013Science} & 6.7 MHz & 3 MHz  & 2.5 MHz \\
Tiecke \textit{et.al}\cite{Tiecke2014Nat} & 0.5 GHz & 6 MHz & 25 GHz \\
Mlynek \textit{et.al}\cite{Mlynek2014NatComm} & 3.7 MHz & 0.27 MHz & 43 MHz \\
Hacker \textit{et.al} \cite{Hacker2019NatPh,Hacker2016Nature}  & 7.8 MHz & 3 MHz & 2.5 MHz \\ 
Yang \textit{et.al} \cite{Yang2019PRL,Yang2019Arx}  & 5.5 MHz & 2.6 MHz & 3.7 MHz \\
Hamsen \textit{et.al}\cite{Hamsen2017PRL} & 20 MHz & 3 MHz & 2 MHz \\
Muniz \textit{et.al}\cite{Muniz2020Nature} & 10.9 kHz & 7.5 kHz & 153 kHz \\
 \hline
\end{tabular}
\label{AtomCavityCouplingStrengths}
\end{center}
\end{table*}

A cavity with a two-level atom (refer Fig. \ref{AtomCavity}) provides an exceptional setting for understanding light-matter interaction \cite{Rabi1936PR,Rabi1937PR,Jaynes1963ProcIEEE}.  The Hamiltonian that describes the interaction of an atom with a single-mode cavity field is \cite{Rabi1936PR,Rabi1937PR}
\begin{align}\label{RabiH}
\hat H_R=\frac{1}{2}\hbar\omega_0 \hat\sigma_z+\hbar \omega \hat a^\dagger \hat a+\hbar g(\hat \sigma_+ +\hat \sigma_-)(\hat a+\hat a^\dagger),
\end{align}
called the Rabi Hamiltonian. Here, $\hat\sigma_z=\ket{e}\bra{e}-\ket{g}\bra{g}$ is the atomic energy operator, $\hat\sigma_+=\ket{e}\bra{g}$ and $\hat\sigma_-=\ket{g}\bra{e}$ are respectively the raising and lowering operators of the atom. $g$ is the atom-field coupling strength.  The states $\ket{e}$ and $\ket{g}$ represent the excited and ground states of the atom respectively.  This Hamiltonian contains the fast-oscillating terms $\hat\sigma_+ \hat a^\dagger$ and $\hat\sigma_- \hat a$. In the interaction picture, these terms exhibit a very fast oscillation $\sim e^{i(\omega_0+\omega)t}$ as compared to $\hat\sigma_+ \hat a$ and $\hat\sigma_- \hat a^\dagger$ which oscillate like $\sim e^{i(\omega_0-\omega)t}$. Hence, under resonance ($\omega_0=\omega$) and small coupling strength $(g<<\omega,\omega_0)$, the fast-oscillating terms can be neglected from the Hamiltonian. This approximation is called rotating-wave approximation (RWA). Therefore, the Hamiltonian, given in Eqn. (\ref{RabiH}), in the weak coupling limit  becomes 
\begin{align}
\hat H_{JC}=\frac{1}{2}\hbar\omega_0 \hat\sigma_z+\hbar \omega \hat a^\dagger \hat a+\hbar g(\hat \sigma_+ \hat a+\hat \sigma_- \hat a^\dagger),
\end{align}
called the Jaynes-Cummings Hamiltonian \cite{Jaynes1963ProcIEEE}. On resonance, that is, $\omega_0=\omega$ and  for a given photon number $n$, the energy eigenvalues of $\hat H_{JC}$ are
\begin{align}\label{dressedeigenvalues}
E_{\pm}(n)=\left(n+\frac{1}{2}\right)\hbar\omega\pm \hbar g\sqrt{(n+1)}, 
\end{align}
 and their corresponding eigenvectors are 
\begin{align}\label{dressedstate}
\ket{\Psi_n^\pm}=\frac{1}{\sqrt{2}}(\ket{e,n}\pm\ket{g,n+1}). 
\end{align}
There are many experimental observations of the emergence of these states  in a coupled atom-cavity system \cite{Raithmaier2004Nat,Yoshie2004Nat,Peter2005PRL, Hennessy2007Nature}. As can be seen from Eqn. (\ref{dressedeigenvalues}) and Fig. \ref{EnergylevelsJC}, eigenvalues are not equally spaced and they have $\sqrt{n}$ dependent scaling which gives rise to a strong nonlinearity in the system. The $\sqrt{n}$ scaling of the eigenvalues experimentally has been observed \cite{Fink2008Nature}. According to Eqn. (\ref{dressedstate}), if the atom-cavity system is initially in the state $\ket{e,n}$, it oscillates in time between the states $\ket{e,n}$ and $\ket{g,n+1}$ with at frequency $2g\sqrt{(n+1)}$ which is the $n$-photon Rabi frequency \cite{AgarwalBook}. Rabi oscillation of an atom inside a cavity is observed experimentally by several groups \cite{Varcoe2000Nature,Meschede1985PRL,Haroche1991EPL}. If the initial state of the cavity field is a coherent state $\ket{\alpha}$, then the atomic population inversion shows collapses and revivals \cite{Eberly1980PRL,Rempe1987PRL}. The existence of these revivals provides direct evidence for the quantum nature of the field \cite{Shore1993JModOptics,Filipowicz1986PRA,BruneSch1996PRL,
Meunier2005PRL}. Some other interesting phenomena such as the vacuum-Rabi splitting \cite{Agarwal1984PRL,Agarwal1985JOSAB,Raizen1989PRL,Hood1998PRL,
Thompson1992PRL,Alsing1992PRA,
Boca2004PRL}, photon blockade \cite{Birnbaum2005Nature, Tang2015ScRep,Deng2017OptExp,Hamsen2017PRL,Huang2018PRL,
Li2019AppSc,Rebic}, photon bunching and anti-bunching \cite{Rein,Hennrich2005PRL, Neuzner2016NatPhoton, Kimble1977PRL,Carmichael1986PRL, Rein,Hennrich2005PRL,Gies2015PRA}, sub-Poissonian and super-Poissonian photon statistics \cite{Walther2006RepProgPhys,Davidovich1996RevModPhys,Rempe1990PRL}, etc could be realized by driving the atom-cavity system. Such system allows to engineer various quantum states of the electromagnetic field in a cavity \cite{Haroche1991PhysWorld,Haroche1995AIPCP,Kimble1998PhysScripta,
Kimble1994APIConfProc,Filipowicz1986JOSAB,Slosser1989PRL,Weidinger1999PRL,
Kuhn1999AppPhyB,Franca2001PRL,Keller2004Nature,Cosacchi2020PRR,
Krastanov2015PRA,Brown2003PRA, Meystre1982PLA,Monteiro2005PhyA,Lutterbach2000PRA,Werlang2008PRA,
Domokos1994PRA,Brune1996PRL,Szabo1996PRA,VillasBoas2003PRA,
Zeng1998PhyLettA,Plastina1999EPJD,Meher2018QInP,Baseia2001CPL}. However, some of the aforementioned phenomena require ultra-strong coupling between the atom and cavity \cite{Frisk2019NatRevPhys}. Reported experimental values of atom-cavity coupling strength are given in Table.~\ref{AtomCavityCouplingStrengths}. For more information on cavity QED,  many excellent reviews are available  \cite{Reiserer2015RevModPhys,Walther2006RepProgPhys,Raimond2001RevModPhys}.


Another regime of coupling that is important in cavity QED is dispersive coupling in which the non-resonant, rather than the resonant, interactions with the field are required. In the dispersive limit, \textit{i.e.}, $(\omega_0-\omega) \gg g$, the Jaynes-Cummings Hamiltonian becomes \cite{Brune1992PRA}
\begin{align}
\hat H_{Disp}=\frac{\hbar\omega_0}{2}\hat \sigma_z+\hbar \omega \hat a^\dagger \hat a+\hbar\chi(\hat \sigma_+ \hat \sigma_-+ \hat \sigma_z \hat a^\dagger \hat a),
\end{align} 
where $\chi=g^2/(\omega_0-\omega)$ is the dispersive coupling strength. The reported experimental values of $\chi$ are given in Table.~\ref{DIspCouplingStrengths}.
Due to the large detuning, the atom and the field do not exchange energy. However, this interaction produces state-dependent shift in the frequency of the atom or cavity  \cite{Brune1992PRA, Vlastakis2013Science}. There are several advantages of non-resonant atom-field interactions than the resonant one. For example, generation of cat states \cite{Brune1992PRA, Vlastakis2013Science} and superposition of number states \cite{Krastanov2015PRA}, nondemolition measurement of photon number \cite{Brune1992PRA}, atomic entanglement measurement \cite{Meher2022AnnderPhys}, realization of quantum phase gate  \cite{Heeres2015PRL, Dong2015PhysLettA}, generation of optical nonlinearity \cite{Imamoglu1997PRL}, control of heat transfer \cite{Meher2020JOSAB}, generation of highly bunched photons \cite{Guo2016PRA}, reading single-qubit states \cite{Bianchetti2009PRA}, demonstration of universal quantum copying machine \cite{Orsag2005JOptB} etc. are possible with non-resonant interaction.  

\begin{table}
\begin{center}
\caption{Reported values of dispersive coupling strengths in experiments}
 \begin{tabular}{|| m{2.5cm} | m{2.5cm} | m{2cm} ||} 
 \hline
References & System & Dispersive coupling strengths $(\chi/2\pi)$  \\ [0.5ex] 
 \hline\hline
Kirchmair \textit{et al.} \cite{Kirchmair2013Nat} & Superconducting cavity- transmon qubit  & 9.4 MHz \\ 
 \hline
Heeres \textit{et al.} \cite{Heeres2015PRL} & Superconducting cavity- transmon qubit  & 8.28 MHz \\ 
 \hline
Inomata \textit{et al.} \cite{Inomata2012PhysRevB} & waveguide resonator-superconducting flux qubit & 40 MHz \\
  \hline
Mallet \textit{et al.} \cite{Mallet2009NatPhys} & coplanar resonator-transmon qubit & 2.175 MHz \\
\hline
Schuster \textit{et al.} \cite{Schuster2007Nature} & coplanar waveguide cavity- Cooper pair box & 8.5 MHz \\
\hline
Kono \textit{et al.} \cite{Kono2018NatPhys} & 3D superconducting cavity- transmon qubit & 1.50 MHz \\
 \hline
Vlastakis \textit{et al.} \cite{Vlastakis2013Science} & waveguide cavity resonator- transmon qubit & 2.4 MHz \\
 \hline
Mirhosseini \textit{et al.} \cite{Mirhosseini2019Nature} & waveguide cavity resonator- transmon qubit & 2.05 MHz \\
 \hline
\end{tabular}
\label{DIspCouplingStrengths}
\end{center}
\end{table}

\subsection{Kerr medium inside a cavity}
\begin{figure}
\begin{center}
\includegraphics[height=6.6cm,width=8cm]{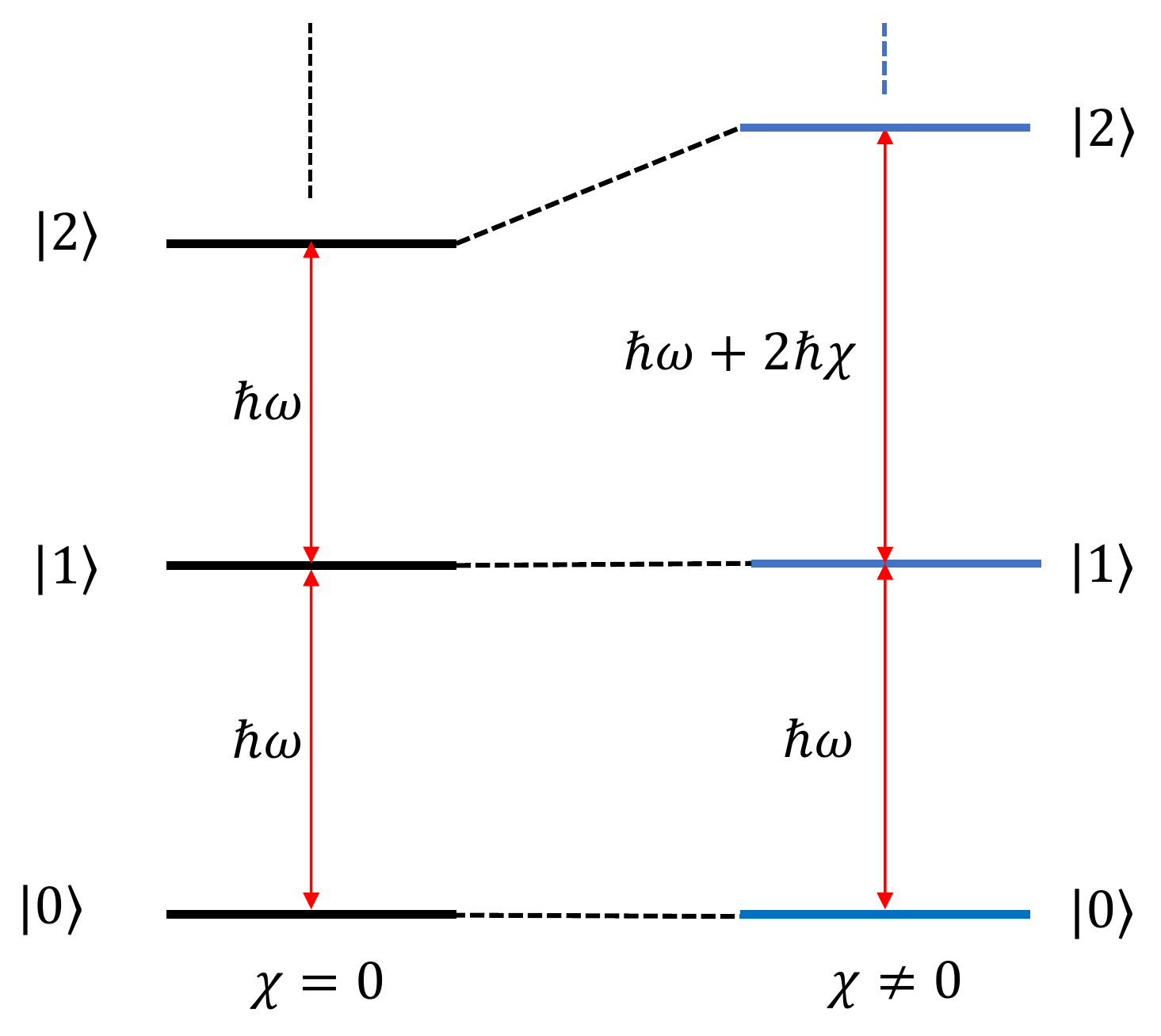}
\caption{Few lowest energy levels of the field in a Kerr cavity for $\chi=0$ (left) and $\chi\neq 0$ (right).}
\label{EnergylevelsKerr}
\end{center}
\end{figure}
Energy levels of the quantized electromagnetic field in an empty cavity are equispaced. Unequal differences between successive levels or anharmonicity  arise on incorporating a nonlinear medium, especially, Kerr medium inside the cavity. The cavity with Kerr nonlinearity is referred as \textquoteleft Kerr cavity\textquoteright{} in subsequent discussions.
Consider a driven cavity containing a nonlinear dispersive medium. Polarization of the medium is \cite{Boyd}
\begin{align}
P=\chi^{(1)}{E}+\chi^{(2)}{E}{E}+\chi^{(3)}{E}{E}{E}+\cdots ,
\end{align}
where $\chi^{(n)}$ is $(n+1)$th rank susceptibility tensor. The energy of the electromagnetic field inside the cavity is
\begin{align}\label{classicalnonlinear}
H=\int_{V}d^3r\frac{1}{2}(\vec{D}\cdot\vec{E}+\vec{H}\cdot\vec{B)},
\end{align}
were $V$ is the modal volume, $\vec{D}=\epsilon_0\vec{E}+\vec{P}$ and $\vec{H}=1/\mu_0 \vec{B}$.
If the field is propagating along $z$-direction and polarization is along the $x$-direction, then $\vec{E}=(E(z,t),0,0),\vec{B}=(0,B(z,t),0)$, $\vec{H}=(0,1/\mu_0 B(z,t),0)$ and the electric flux density $\vec{D}=(\epsilon_0 E+P,0,0)$. If the medium is centro-symmetric and the driving is intense then the third-order nonlinearity $\chi^{(3)}$ is larger than the linear susceptibility $\chi^{(1)}$ and second-order susceptibility $\chi^{(2)}$. With this assumption, the Hamiltonian given in Eqn. (\ref{classicalnonlinear}) can be written as \cite{Hiroo2008JPhyDAppPhy}
\begin{align}
H=\int_V d^3r \frac{1}{2}\left[\left(\epsilon E^2+\frac{1}{\mu_0}B^2\right)+\chi^{(3)}E^4\right].
\end{align}
Under RWA, the above Hamiltonian becomes \cite{Drummond1980JPhyAMatGen}
\begin{align}
\hat H=\hbar\omega \hat{a}^\dagger \hat{a}+\hbar \chi \hat{a}^{\dagger 2}\hat{a}^2,
\end{align}
where 
\begin{align}
\chi=\frac{3\hbar \omega^2}{8\epsilon_0^2}\int \chi^{(3)}\vert u(r)\vert^4 d^3 r=\frac{3\hbar \omega^2 \chi^{(3)}}{4\epsilon_0 V_{eff}\epsilon_r^2},
\end{align}
is the strength of Kerr nonlinearity. The mode function $u(r)$ satisfies $\int [u^*(r)(1+\chi^{(3)}/\epsilon)u(r)]d^3r=1$. Experimentally reported values of $\chi^{(3)}$ and $\chi$ are listed in the Tables. \ref{NonlinearStrength} and \ref{NonlinearStrengthX}. Few more values of $\chi^{(3)}$ of various materials can be found in the Ref. \cite{Ferretti2012PRB,Youngblood2017ACSPh}.
\begin{table}[t]
\begin{center}
\caption{Reported values of Kerr coefficients $\chi^{(3)}$.}
 \begin{tabular}{|| m{3.5cm}|m{2.5cm} | m{2.1cm}||} 
 \hline
References & System & Nonlinear strength ($m^2/V^2$)  \\ [0.5ex] 
 \hline\hline
 Fushman \textit{et al.} \cite{Fushman2008Science}& InAs quantum dot in photonic crystal & $2.4\times  10^{-10}$  \\
  \hline
Wang \textit{et al.} \cite{Wang2021NatComm}&  J-aggregate cyanine molecules in Fabry-Perot cavity & $0.747\times 10^{-17}$   \\
 \hline
Woodward \textit{et al.} \cite{Woodward20162DMat}& monolayer MoS$_2$ & $2.4 \times 10^{-19}$    \\
 \hline
Youngblood \textit{et al.} \cite{Youngblood2017ACSPh} & Black Phosphorus & $1.4\times 10^{-19}$   \\
 \hline
\end{tabular}
\label{NonlinearStrength}
\end{center}
\end{table}

\begin{table}[t]
\caption{Reported values of Kerr nonlinearity strength ($\chi$).}
\begin{center}
 \begin{tabular}{|| m{3.3cm}|m{2.5cm} | m{2cm}||} 
 \hline
 References& System & Nonlinear strength ($\chi/2\pi$)  \\ [0.5ex] 
 \hline\hline
Kirchmair \textit{et al.} \cite{Kirchmair2013Nat}&Superconducting cavity- transmon qubit & 0.32 MHz  \\
  \hline
Heeres \textit{et al.} \cite{Heeres2015PRL}& Superconducting cavity- transmon qubit &0.1 MHz    \\
 \hline
 Vrajitoarea \textit{et al.} \cite{Vrajitoarea2020NatPhys}&Superconducting microwave cavities& 12.5 MHz    \\
 \hline
\end{tabular}
\label{NonlinearStrengthX}
\end{center}
\end{table}
The energy of $n$ photons in the presence of Kerr nonlinearity is $\bra{n}H\ket{n}=n\hbar\omega+n(n-1)\hbar \chi$. As can be seen from the Fig. \ref{EnergylevelsKerr}, the energy of the electromagnetic field in a Kerr cavity is proportional to the square of number of photons and the energy level becomes anharmonic \cite{Imamoglu1997PRL,Rebic1999JOptB,Auffeves2007PRA,
Fushman2008Science}. Anharmonicity in the energy levels results in strong photon-photon interaction that gives rise to various interesting phenomena such as the photon blockade \cite{Imamoglu1997PRL,Grangier1998PRL,Birnbaum2005Nature,
Tang2015ScRep,Shen2015PRA,Miranowicz2014PRA,
Miranowicz2016PRA,Andrew2021ScAdv}, bunching and antibunching of photons \cite{Zhang2016ScRep,Gies2015PRA}, slow light propagation \cite{Dey2007PRA}, etc. The anharmonicity in the energy levels is used for demonstrating quantum gates \cite{Hiroo2008JPhyDAppPhy,Meher2019JPhysB}, engineering quantum states \cite{Puri2017npjQInf}, processing quantum information \cite{Mabuchi2002Science} and the realization of strongly correlated states of light and matter \cite{Chang2014NatPhot,Kamanasish2017NJP},  etc.   

\subsection{Circuit Quantum Electrodynamics (CQED)} 
\begin{figure}
\begin{center}
\includegraphics[height=6.4cm,width=8cm]{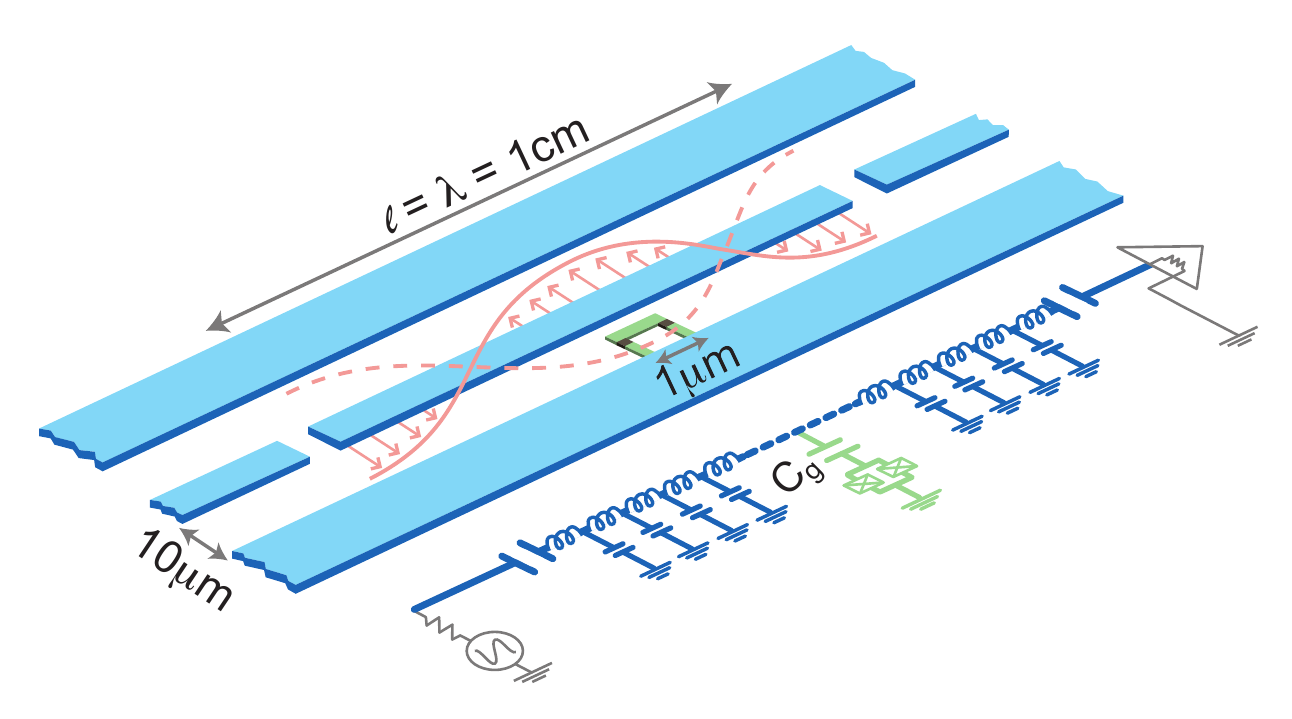}
\caption{Schematic of superconducting coplanar waveguide capacitively coupled to a Cooper-pair box qubit. The figure is reproduced from the Ref. \cite{Blais2004PRA} with permission from American Physical Society.}
\label{SuperconductingCavity}
\end{center}
\end{figure}
A brief mention  of circuit quantum electrodynamics (CQED) is apt here as there are  many commonalities between CQED and cavity QED.  Some obvious advantages are the possibility of tunability and higher strength of qubit-field interaction compared to atom-cavity systems \cite{Blais2020NatPhys}.  In its basic form, a circuit QED configuration (Fig. \ref{SuperconductingCavity}) is a Cooper-pair box (CPB) enclosed in a coplanar waveguide resonator that supports a microwave field.  Here, CPB is the \textquotedblleft artificial atom\textquotedblright, as it plays the role analogous to that of the atom in a cavity QED setup. The Hamiltonian for the system is \cite{Blais2021RevModPhys}
\begin{align}\label{H_cQED_1} 
\hat H_{CQED}=4E_c\hat{N}^2-E_J\cos\hat{\phi}-\hbar\omega\hat{a}^\dagger\hat{a}.
\end{align}
The first term corresponds to the square of the number of charges in CPB.  This term arises as the energy associated with a capacitor is proportional to the square of the charge on the capacitor.   The second term whose strength is $E_J$ is the Josephson junction energy  and $\hat\phi$ is the operator corresponding to the phase difference across the junction.  The last term in the Hamiltonian is to account for field mode. 

Writing the operators $\hat\phi$ and $\hat{N}$ in terms of creation and annihilation operators
 \begin{eqnarray}
			\hat\phi&=&\left(\frac{2E_c}{E_J}\right)\left(\hat{b}^\dagger+\hat{b}\right),\\
			\hat N &=&\frac{i}{2}\left(\frac{E_J}{2E_c}\right)\left(\hat{b}^\dagger-\hat{b}\right),
\end{eqnarray}
and invoking rotating-wave approximation, 
one can recast Eqn. (\ref{H_cQED_1}) to \cite{Krantz2019AppPhysRev}
 \begin{equation}\label{H_cQED2}
 	\hat H_{CQED}\approx\hbar\omega \hat{a}^\dagger \hat{a}+\hbar\omega_c \hat{b}^\dagger \hat{b}-\frac{E_c}{2}\hat{b}^\dagger \hat{b}^\dagger \hat{b}\hat{b}+\hbar g(\hat{b}^\dagger \hat{a}+\hat{b}\hat{a}^\dagger).
 \end{equation} 
This Hamiltonian is analogous to the Hamiltonian for two coupled cavities in which one of the cavities contains a Kerr medium. 

To make the above Hamiltonian analogy with cavity QED Hamiltonian, we need to restrict the description of the transmon to its first two levels. Hence, replacing $\hat b \rightarrow \hat\sigma_-=\ket{g}\bra{e}$ and $\hat b^\dagger \rightarrow \sigma_+=\ket{e}\bra{g}$, we get the Jaynes-Cummings Hamiltonian \cite{Krantz2019AppPhysRev}
\begin{align}
\hat H_{JC}=\frac{1}{2}\hbar\omega_q \hat\sigma_z+\hbar \omega \hat a^\dagger \hat a+\hbar g(\hat \sigma_+ \hat a+\hat \sigma_- \hat a^\dagger),
\end{align}
where $g$ is the coupling strength between the oscillator and transmon. Detailed review on circuit QED can be found in Refs.   \cite{Gu2017PhysRep,Krantz2019AppPhysRev,Blais2020NatPhys,Blais2021RevModPhys}.
\section{Two coupled cavities} 
\begin{figure}
\begin{center}
\includegraphics[height=4.5cm,width=8cm]{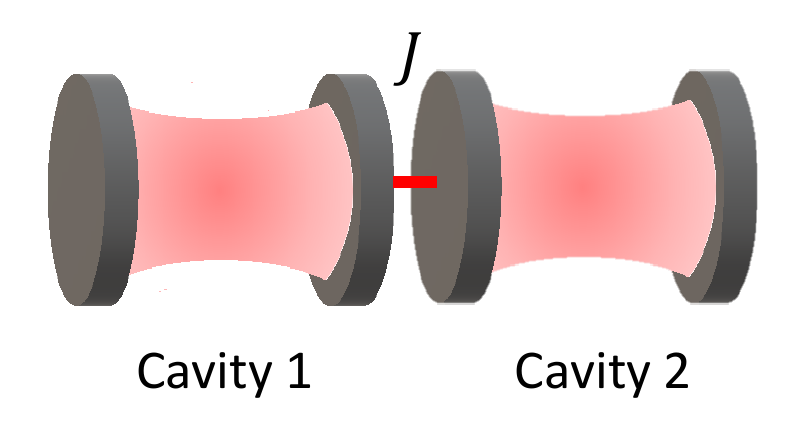}
\caption{Two cavities are arranged in close proximity so as to exchange photons among them is possible. The coupling strengths is $J$. }
\label{Coupledcavities}
\end{center}
\end{figure} 
Technological progress in the fabrication of high-$Q$ cavities has rendered it possible to couple several cavities to build an extended quantum network \cite{Notomi2008NatPhy,Akahane2003Nature, Majumdar2012PRA,Cho2008PRL,Zhukovsky2007PRL}. The coupling between two distant cavities  is important for building photonic integrated circuits \cite{Sato2011Nature}. A simple coupled system is a two coupled cavities \cite{Liew2010PRL,Bose2014NatPhotonics,Kapfinger2015NatComm,
Hamel2015NatPhot,Bayindir2000PRL,Zhao2020OptExp}, as can be seen in Fig. \ref{Coupledcavities}. The two cavities must be placed very close to each other (within a few wavelengths) such that the evanescent fields of photons of both the cavities overlap \cite{Ilchenko1994OptComm,MajumdarRund2012PRB}. The rate of exchange of energy depends on the coupling strength $(J)$ which, in turn, depends on the overlap of the spatial profiles of resonant modes \cite{Hartmann2016JOpt,Bayer1998PRL}. The interaction Hamiltonian for the cavities coupled \textit{via} evanescent waves is \cite{Irvine2006PRL,EbelingSpringer}
\begin{align}\label{interctionHamiltonian}
H_{int}=\int\epsilon E_1E_2^* dV,
\end{align}
where $\epsilon$ is the relative permittivity profile for the coupled cavities, and $E_1$ and $E_2$ represent the electric fields of these cavities. The range of integration extends over the volume of the coupled cavities. The two cavities are considered to be non-ideal in a sense their electric field distributions extend beyond the cavity boundaries. If the cavities are ideal then the electric fields are confined within the respective cavities and the interaction energy vanishes. Using the expression of single-mode electric field of the individual cavities
$\hat{E}_j=\sqrt{\frac{\hbar\omega_j}{2\epsilon_0\epsilon_{rj}V_j}}(\hat a_j+\hat a_j^\dagger)u_j(z),$ the interaction Hamiltonian given in Eqn. (\ref{interctionHamiltonian}) can be written as
\begin{align}
\hat H_{int}= &\sqrt{\frac{\hbar^2\omega_1\omega_2\epsilon^2}{4\epsilon_0^2\epsilon_{r1}\epsilon_{r2}V_1V_2}}\int    (\hat{a}_1+\hat{a}_1^\dagger)(\hat{a}_2+\hat{a}_2^\dagger) u_1(z)u_2^*(z) dV,\nonumber\\
= & \hbar J(\hat a_1+\hat a_1^\dagger)(\hat a_2+\hat a_2^\dagger),
\end{align}
where 
\begin{align}\label{CouplingStrength}
J=\sqrt{\frac{\omega_1}{2\epsilon_0\epsilon_{r1}V_1}}\sqrt{\frac{\omega_2}{2\epsilon_0\epsilon_{r2}V_2}}\int  \epsilon u_1(z)u_2^*(z) dV, 
\end{align}
is the coupling strength between the cavities. 
Here, $u_j(z)$ is the mode function of the $j$th cavity in the absence of other cavity, $\epsilon_{rj}$ is the relative permittivity of the medium present and $V_j$ is the mode volume  of the $j$th cavity respectively. Reported experimentally achievable values of $J$ in various coupled-cavity configurations are given in Table. \ref{CouplingStrengths}.  

\begin{table*}
\caption{Reported experimental values of parameters in coupled cavities}
\begin{center}
 \begin{tabular}{|| c | m{2.5cm} | m{2.5cm} | m{2.5cm}  ||} 
 \hline
References &System & Resonance frequencies (Hz) & Coupling strengths (Hz) \\ [0.5ex] 
 \hline\hline
Sato \textit{et al.} \cite{Sato2011Nature} & Silicon-based photonic crystal cavities & 1.93$\times 10^{14}$ & 5.8 $\times 10^{10}$  \\ 
 \hline
Majumdar \textit{et al.} \cite{Majumdar2012PRA}& GaAs photonic crystal cavities  & 3.33$\times 10^{14}$ & 1.3$\times 10^{12}$  \\ 
 \hline
Majumdar \textit{et al.}\cite{MajumdarRund2012PRB}&GaAs photonic crystal cavities & 1.169$\times 10^{14}$ & 11$\times 10^{10}$  \\
  \hline
  Cai \textit{et al.}\cite{Cai2013APL}&GaAs photonic crystal cavities & 3.15$\times 10^{14}$ & 18 $\times 10^{9}$ \\
 \hline
 Konoike \textit{et al.}\cite{Konoike2019APLPhot, Konoike2016ScAdv,Konoike2014IEEEPhotConf}&Si photonic crystal nanocavities & 1.9$\times 10^{14}$ & 25 $\times 10^{9}$ \\
 \hline
 Du \textit{et al.}\cite{Du2016ScRep}& Silicon-on-insulator nanobeam photonic molecule& 1.9$\times 10^{14}$ & 1.9$\times 10^{12}$  \\
 \hline
\end{tabular}
\label{CouplingStrengths}
\end{center}
\end{table*}

Coupling between cavities is also possible through inductive coupling, capacitive coupling \cite{Peropadre2013PRB}, waveguide coupling \cite{Sato2011Nature}, etc., and also through various other mechanisms \cite{Cai2013APL,Ilchenko1994OptComm, Majumdar2012PRA, Du2016ScRep}. However, to build a large-scale array, coupling between two distant cavities is essential which is possible through the waveguide coupling. But to realize strong coupling between the cavities, it is necessary to choose the cavity and waveguide parameters such that they satisfy mode-mismatch conditions
$\delta_{in} \ll \Delta_{FP}$ and $
2\theta_c \approx (2m+1)\pi$ \cite{Sato2011Nature}, 
where $\delta_{in}$ is the coupling bandwidth which decides the decay rate of photons from the cavity to the waveguide, $\Delta_{FP}$ is the free spectral range and $\theta_c$ represents the propagation phase of the waveguide. When these two conditions are satisfied, photons from the cavity hardly leak out to the waveguide. But the two cavities are still be coupled indirectly through a forced oscillation of the waveguide modes \cite{Sato2011Nature}. Essentially, photons from one cavity flow to the other cavity without populating the waveguide mode. Recently, strong coupling between two cavities has been demonstrated experimentally even though the distance between them exceeded 100 wavelengths \cite{Sato2011Nature}.

The total Hamiltonian for the coupled cavities is
\begin{align}\label{Ch1HWRWA}
\hat H&=\hbar\omega_1 \hat a_1^\dagger \hat a_1+\hbar\omega_2 \hat a_2^\dagger \hat a_2+\hbar J(\hat a_1+\hat a_1^\dagger)(\hat a_2+\hat a_2^\dagger),
\end{align}
where $\omega_1$ and $\omega_2$ are the resonance frequencies of the first and second cavities respectively. Under RWA, the above Hamiltonian reduces to \cite{Estes1968PR,Castro2002JOptBQS}
\begin{equation}\label{Ch1HRWA}
\hat H=\hbar \omega_1\hat{a}_1^\dagger \hat{a}_1+\hbar\omega_2 \hat{a}_2^\dagger \hat{a}_2+ \hbar J(\hat{a}_1^\dagger \hat{a}_2+ \hat{a}_1 \hat{a}_2^\dagger).
\end{equation}
For this Hamiltonian, the excitation number operator $\hat N=\hat a_1^\dagger \hat a_1+\hat a_2^\dagger \hat a_2$ is a conserved quantity, \textit{i.e.}, $[\hat H,\hat N]=0$. The unitary dynamics due to $\hat H$ is restricted to respective invariant subspaces corresponding to the photon numbers.

Under resonance condition $(\omega_1=\omega_2=\omega)$, the equations of motion of annihilation operators of the respective cavities are
\begin{align}
\dot{\hat{a}}_1=-i\omega \hat{a}_1-iJ\hat{a}_2,\\
\dot{\hat{a}}_2=-i\omega \hat{a}_2-iJ\hat{a}_1.
\end{align}
By solving these two coupled differential equations, we get the time dynamics of the number operators for individual cavities to be
\begin{align}
\langle \hat{a}_1^\dagger \hat{a}_1 (t)\rangle   =\cos Jt \langle \hat{a}_1^\dagger \hat{a}_1 (0)\rangle   -i\sin Jt \langle \hat{a}_2^\dagger \hat{a}_2(0)\rangle, \\
\langle \hat{a}_2^\dagger \hat{a}_2 (t)\rangle   =\cos Jt \langle \hat{a}_2^\dagger \hat{a}_2 (0)\rangle   -i\sin Jt \langle \hat{a}_1^\dagger \hat{a}_1(0)\rangle.  
\end{align}
Hence, a field whose average number of photons is $n$ initially in the first cavity will evolve so that the photons are completely transferred to the second cavity at time $t=\pi/2J$ and again return to the first cavity at time $t=\pi/J$. The oscillation of the field between the cavities is called Rabi oscillation. For coupled cavities, the Rabi oscillation frequency is $2J$.      
Recently, Rabi oscillation with a period of 54 ps between two photonic crystal cavities has been observed experimentally \cite{Sato2011Nature}. 

If the cavities are not resonant ($\omega_1\neq \omega_2$), then the complete transfer of the photons does not occur. For large detuning ($\Delta=\omega_1- \omega_2 \gg J$), all the photons will get localized in its initial cavity and the transfer of photons to the other cavity does not happen \cite{Sato2011Nature}. Hence, detuning allows controlled transfer of photons between the cavities, which is experimentally demonstrated by Sato \textit{et al.} \cite{Sato2011Nature}. Thus, the tuning of resonance frequencies enables remote control of photon transfer. There exist many ways of tuning the resonance frequencies such as nanofluidic tuning \cite{Vignolini2010ApplPhysLett,Vignolini2010AppPhysLett2}, nonlinear optical tuning \cite{Fushman2007APL}, thermo-optic tuning \cite{DUndar2012APL}, nano-mechanical tuning \cite{Hopman2006OptExp}, photochromic tuning \cite{Cai2013APL}, etc. to mention a few. However, controlled transfer photons between two non-resonant cavities is possible by suitably choosing the values of detuning and coupling strength $J$ or including Kerr nonlinearity \cite{Meher2016JOSAB,Meher2017ScRep}. And, recently, Konoike \textit{et al.} have been experimentally demonstrated the adiabatic transfer of photons between two non-resonant cavities by adding one intermediate cavity \cite{Konoike2016ScAdv, Konoike2013PRB}. 

\section{Coupled cavity array}
The Hamiltonian given in Eqn. (\ref{Ch1HRWA}) can be extended to describe an array with a large number of cavities. Consider a system $N$ linearly coupled cavities. The Hamiltonian for the system is ($\hbar=1$)
\begin{equation}\label{Ch2arrays}
\hat{H}=\sum_{l=1}^N {\omega}_l \hat{a}_l^\dagger \hat{a}_l +\sum_{l=1}^{N-1} 
{J}_l(\hat{a}_l^\dagger \hat{a}_{l+1}+\hat{a}_l \hat{a}_{l+1}^\dagger),
\end{equation}
where $\hat{a}_l$ and $\hat{a}_l^{\dagger}$ are respectively the  annihilation and creation operators, and ${\omega}_l$ is the resonance frequency of the $l$-th cavity.  The coupling strength between the $l$-th and $(l+1)$-th cavities is ${J}_l$.
\subsection{Transfer of photons}
Perfect transfer of photons in a cavity array is essential for quantum state transfer and entanglement generation. In this subsection, we discuss the possibility of perfect transfer of photons through the array. The transfer is said to be perfect if the probability of transfer to the target cavity is unity. 

\begin{figure}
\centering
\includegraphics[width=9cm,height=5cm]{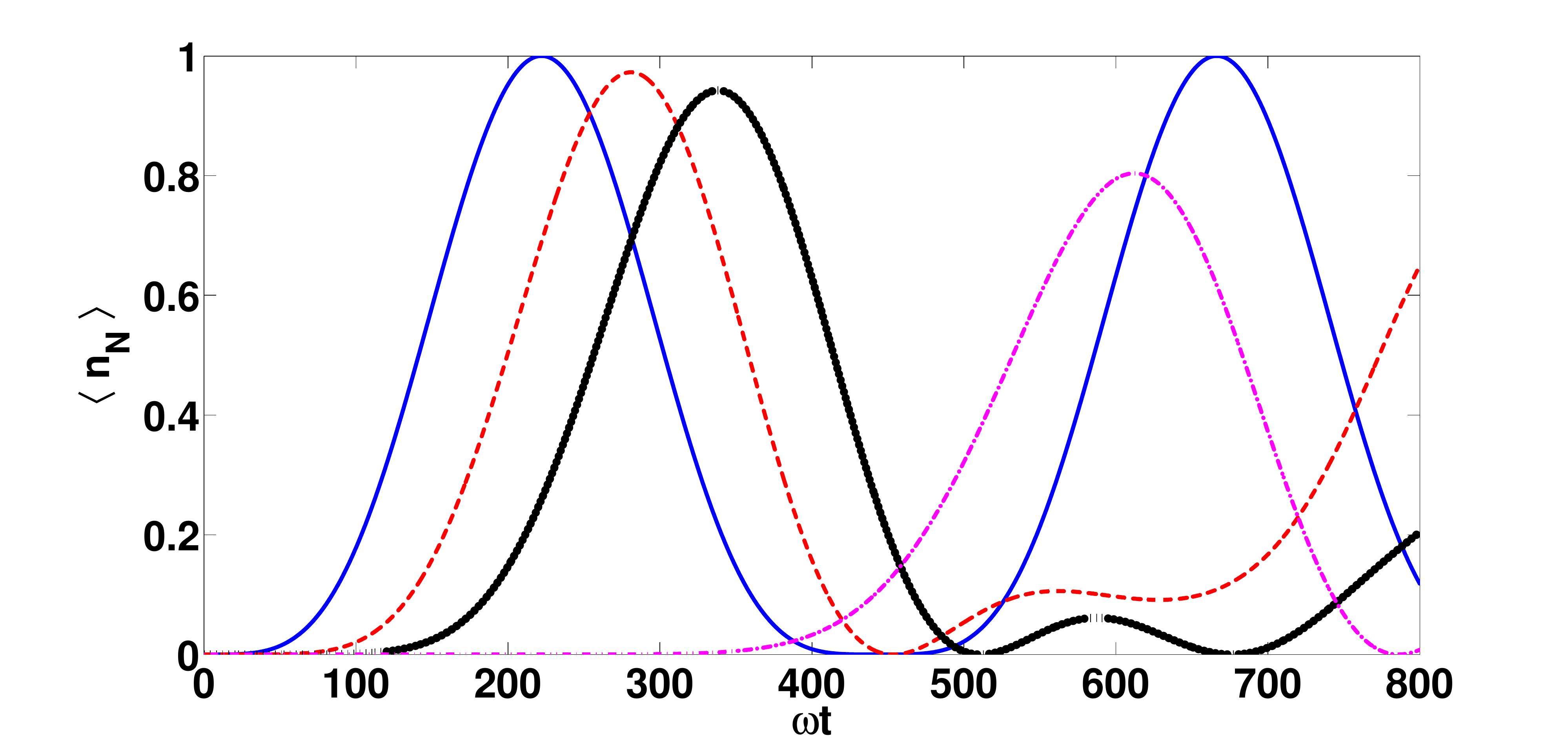}
\caption{Average number of photon in the end cavity as a function
 of $\omega t$ in a resonantly coupled cavities with equal coupling strengths. The array contains $N$ number of cavities with $N=$3 (solid line), 4 (dashed), 5 (dotted) and 10 (dot-dashed). We set $J/\omega=0.01$. }
\label{Ch2uniformCoupling}
\end{figure}

Consider the simplest situation where all the coupling strengths ${J}_l$  are equal $(=J)$ and all the resonance frequencies are equal $(=\omega)$, then the Hamiltonian given in Eqn. (\ref{Ch2arrays}) becomes
\begin{align}\label{HomogeneousArray}
\hat{H}={\omega}\sum_{l=1}^N  \hat{a}_l^\dagger \hat{a}_l +{J}\sum_{l=1}^{N-1} 
(\hat{a}_l^\dagger \hat{a}_{l+1}+\hat{a}_l \hat{a}_{l+1}^\dagger).
\end{align}
This Hamiltonian conserves the number of excitations as $[\hat{H},\sum_{l=1}^N \hat a_l^\dagger \hat a_l]=0$ and hence, the Hamiltonian can be diagonalized in the subspace corresponding to a given excitation number. Defining normal mode operators for the cavity array 
\begin{align}
\hat{c}_k(t)=\sum_{j=1}^N \hat a_j(t)S(j,k),
\end{align}
the inverse transformation is
\begin{align}\label{Ch2inversetrans}
\hat a_j(t)=\sum_{k=1}^N \hat c_k(t)S(j,k).
\end{align}
The transformation matrix  $S(j,k)$ is
\begin{align}
S(j,k)=\sqrt{\frac{2}{N+1}}\sin\left(\frac{j\pi k}{N+1}\right).
\end{align}
Using the orthogonality relation 
\begin{align}
\sum_{j=1}^N\sin\left(\frac{j\pi k}{N+1}\right)\sin\left(\frac{j\pi m}{N+1}\right)=\frac{N+1}{2}\delta_{km},
\end{align}
we write the Hamiltonian given in Eqn. (\ref{HomogeneousArray}) to be
\begin{align}\label{Ch2HinNormalMode}
\hat H=\sum_{k=1}^N \Omega_k\hat c_k^\dagger \hat c_k,
\end{align}
where $\hat{c}_k$ and $\hat{c}_k^\dagger$ are the creation and annihilation operators for the $k$-th normal mode.
The Hamiltonian in the normal mode coordinates corresponds to a collection of independent oscillators. The normal mode frequencies are  \cite{Notomi2008NatPhy}
 \begin{align}\label{Ch2normalmodefrequency}
 \Omega_k=\left(\omega+2J\cos \frac{\pi k}{N+1}\right),~~~~ k=1,2,3,.... ,N.
\end{align} 
The evolution equation for $k$-th normal mode operator is
\begin{align}
&\frac{d}{dt}\hat c_k=i[\hat{H},\hat c_k]=i\left[\sum_{n=1}^N \Omega_n\hat c_n^\dagger \hat c_n,\hat c_k\right]=-i\Omega_k \hat c_k.
\end{align}
The solution of the above equation is
\begin{align}
\hat c_k(t)=e^{-i\Omega_k t}\hat c_k(0).
\end{align}
Using the inverse transformation given in Eqn. (\ref{Ch2inversetrans}),
the annihilation operator for the $j$-th mode is
\begin{align}
\hat a_j(t)=\sum_{l=1}^N \sum_{k=1}^N e^{-i\Omega_k t} \hat a_l(0) S(l,k) S(j,k).
\end{align}
 The average number of photons in the $j$-th cavity at time $t$ is given by
 \begin{equation}
\langle n_j(t)\rangle=\langle \hat a_j^\dagger \hat a_j(t)\rangle=
\sum_{l=1}^N \vert G_{jl}\vert^2 \langle \hat a_l^\dagger \hat a_l(0)\rangle,
\end{equation}
where
\begin{align}\label{Ch2Gjl}
G_{jl}=&\frac{2}{N+1}
\sum_{k=1}^N e^{-i\Omega_k t}\sin\left(\frac{j\pi k}{N+1}\right)\sin\left(\frac{l\pi k}{N+1}\right).
\end{align}  
Consider a single photon in the first cavity, that is,  
$\langle \hat a_l^\dagger \hat a_l(0)\rangle=\delta_{1,l}$. Then, the average photon number in the last cavity is
 \begin{align}\label{averagephotonhomogeneous}
\langle n_N(t)\rangle & =\langle \hat a_N^\dagger \hat a_N(t)\rangle=\vert G_{N1}\vert^2\nonumber\\
&=\left\vert\frac{2}{N+1}
\sum_{k=1}^N e^{-i\Omega_k t}\sin\left(\frac{N\pi k}{N+1}\right)\sin\left(\frac{\pi k}{N+1}\right)\right\vert^2.
\end{align}
Time evolution $\langle n_N\rangle$ for arrays with $N=$3, 4, 5 and 10 cavities 
 respectively are shown in Fig. \ref{Ch2uniformCoupling}.   From the figure
 it is clear that complete transfer, that is, max[$\langle n_N(t)\rangle]=1$ occurs if the array has three cavities \cite{Bose2003PRL,Christandl2004PRL,Godsil2012PRL,Meher2017ScRep,
 Felicetti2014PRA}. 
 Maximum of $\vert G_{N1}\vert^2$ decreases with increasing number of cavities. This is inferred from Eqn.~(\ref{Ch2Gjl}) on noting that for large $N$, 
$\sin(Nk\pi/N+1)\approx \sin(k\pi)=0$ and $G_{N1}$ tends to zero. What happens in the limit of large $N$ is that during the time evolution a single photon is shared by all the cavities. Hence, detecting the photon in any of the cavities with unit probability is not possible.
\begin{figure}
\centering
\includegraphics[width=9cm,height=5cm]{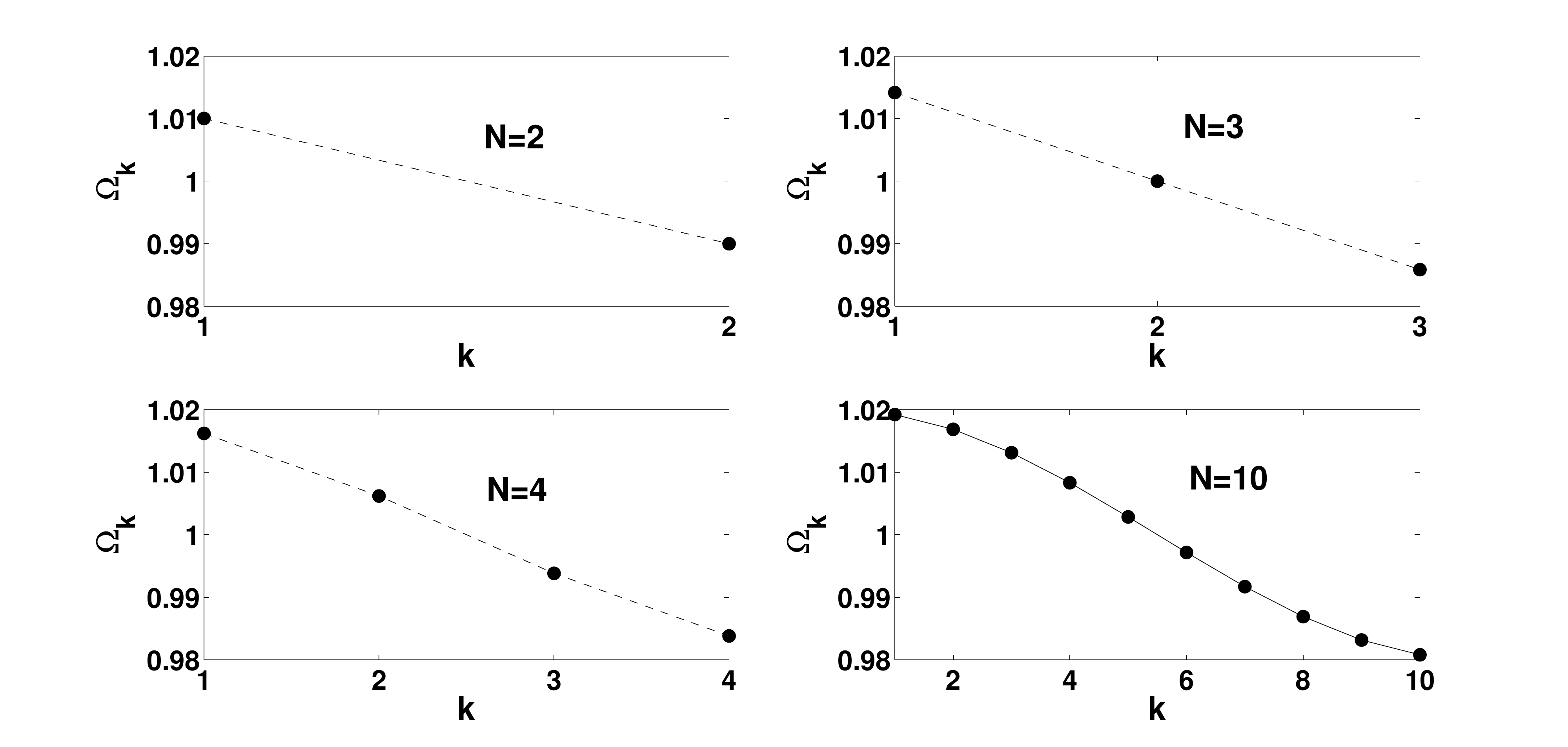}
\caption{Normal mode frequencies for $N=2,3,4$ and 10. We set $J/\omega=0.01$}
\label{NormalModeFrequencies23410}
\end{figure}
In order to understand the transfer behavior of the array, the normal mode frequencies $\Omega_k$ given in Eqn. (\ref{Ch2normalmodefrequency})  are plotted in Fig. \ref{NormalModeFrequencies23410}. For two cavities, the normal mode frequencies are $\Omega_1=\omega+J$ and $\Omega_2=\omega-J$ which lie on a straight line. Similarly, the normal mode frequencies for three cavities are $\Omega_1=\omega+\sqrt{2}J$, $\Omega_2=\omega,$ and $\Omega_3=\omega-\sqrt{2}J$. These are also collinear. However, for $N\geq 4$, the dispersion relation becomes nonlinear. Due to this nonlinear dispersion relation, propagation of photons suffers dispersion and the complete transfer does not occur for $N>3$ (refer Fig. \ref{Ch2uniformCoupling}). Thus, homogeneous coupling does not provide perfect transfer of a photon in the array having more than three cavities.

Perfect transfer of photons demands a correct combination of the coupling strengths in the array. A form of site-dependent coupling strengths that provide perfect transfer of photons is \cite{Matthieu2012OptLett,Meher2017ScRep,Perez2013PRA,Perez2013PRAFeb,Chapman2016NatComm,
Joglekar2013EPJAP, YangLiu2015NJP}
\begin{align}\label{sitedependentcoupling}
J_l=\sqrt{l(N-l)}J,
\end{align}  
where $J$ is a constant and $N$ is the number of cavities in the array. This form of coupling strength has been considered in the context of transferring quantum states in spin chain \cite{Christandl2004PRL}.  However, it is also possible to derive the above coupling strengths from a duality relation between \textquotedblleft $N-1$ photons in two coupled cavities\textquotedblright{} and \textquotedblleft single photon in $N$ cavities\textquotedblright{} \cite{Meher2017ScRep}. The Hamiltonian given in Eqn. (\ref{Ch2arrays}) becomes
\begin{align}\label{sitedependentH}
\hat{H}=\sum_{l=1}^N {\omega}_l \hat{a}_l^\dagger \hat{a}_l +\sum_{l=1}^{N-1} 
\sqrt{l(N-l)}J(\hat{a}_l^\dagger \hat{a}_{l+1}+\hat{a}_l \hat{a}_{l+1}^\dagger).
\end{align} 
If the cavities are resonant, the eigenvectors of the Hamiltonian in single-photon subspace are \cite{Meher2017ScRep}
\begin{align}\label{Ch2EigstateHA}
\ket{X_{n+1}}=&\frac{1}{\sqrt{2^{N-1}}}\sum_{k=0}^{N-1-n}\sum_{k'=0}^{n} (-1)^{k'}~^{N-1-n}C_k \nonumber\\
&~~~~~~~~~~~~~~~~~~~~~~~~~\times ~^nC_{k'}\sqrt{\frac{^{N-1}C_n}{^{N-1}C_{r}}}\ket{r+1}\rangle,
\end{align}
with eigenvalues $E_{n+1}=(N-1)\omega+(N-1-2n)J$, where $r=n+k-k'$ and $n=0,1,...,N-1$. Here, the state $\ket{r+1}\rangle$ represents the state to  having a single photon in the $(r+1)$th cavity and other cavities are in their respective vacuua. Now, consider a photon in the first cavity and the other cavities in vacuum as the initial state, then the initial state is $\ket{1}\rangle=\ket{1}\ket{0}...\ket{0}$. The state at later times is
\begin{align}\label{Ch2UnitEvol}
&e^{-i \hat Ht}\ket{1}\rangle=e^{-i \hat Ht}\ket{1}\ket{0}...\ket{0}\nonumber\\
&=e^{-i\omega t}\sum_{k=0}^{N-1} 
\sqrt{\binom{N-1}{k}}(\cos Jt)^{N-1-k}(-i\sin Jt)^{k}\ket{k+1}\rangle.
\end{align}
The probability of finding the photon at the end of the array, that is, in $N$th cavity is
\begin{align}
P=\vert\langle \bra{N}e^{-i \hat Ht}\ket{1}\rangle \vert^2=(\sin^2 Jt)^{N-1}.
\end{align}
It is to be noted that the probability $P$ becomes unity at $t=\pi/2J$, which corresponds to the perfect transfer of the photon from the first cavity to the last cavity. Hence, the choice of coupling strengths given in Eqn. (\ref{sitedependentcoupling}) allows the perfect transfer of a photon through the cavity. This comes from the fact that, in contrast to the homogeneous coupling, the site-dependent couplings make the dispersion relation linear which results in dispersionless transport. The site-dependent couplings not only allow the perfect transfer of a photon between two end cavities, but also allow perfect transfer between two symmetrically located cavities in the array, that is, from $l$th cavity to $(N+1-l)$th cavity \cite{Matthieu2012OptLett,Meher2017ScRep,Perez2013PRA}. Interestingly, transfer of a photon is controllable between any two arbitrary cavities if Kerr nonlinearity is included in the cavities \cite{Meher2017ScRep}.

\subsection{Controllable photon transfer}
Controlled transfer of photons is essential for transferring information between selected nodes in a quantum network. To control the transfer, one may require to tune the resonance frequencies \cite{Liao2010PRA,Duan2020ChPhyB} or coupling strengths \cite{Liu2021ChJourPhys,Cryan2011AdvOptElec} of the array, or embedding material medium such as atoms \cite{Zhou2008PRL,Zhou2008PRA, Zhou2009PRADec,Gong2008PRA,Yan2012PRA,Yan2014PRANov,Zhou2013JOSAB,
Zang2010JPhyB,Lu2010PRA,Chang2011PRA,Hu2007PRAJul,Li2017JOSAB,
Yang2018IJTP,Yan2020QST} or Kerr nonlinearity \cite{Imamoglu1997PRL,Rebic1999JOptB,Miranowicz2006JPhyB,Hiroo2008JPhyDAppPhy,Ferretti2010PRA,
Liew2013NJP,Chang2014NatPhot,Kamanasish2017NJP, Meher2017ScRep,
Biella2015PRA,Meher2016JOSAB,Manosh2018JNOPM,Zhou2020OptExp,Lin2020IJTP} in the array.


Recently, Konoike \textit{et al.} have experimentally demonstrated the controlled transfer of photons from one cavity $(A)$ to another cavity $(B)$ by introducing an additional cavity $(C)$ in the middle \cite{Konoike2016ScAdv}. Cavities were fabricated in a two-dimensional photonic crystal. The coupling strength between the cavities $A$ and $C$ was $\sim$25 GHz and between $B$ and $C$ was $\sim$ 16GHz. If the resonance frequencies of the cavities satisfy $\omega_C<<\omega_A <\omega_B$, then the Rabi oscillation between the cavities $A$ and $B$  is suppressed. Under these conditions, if the photons are initially in cavity $A$, they remain in cavity $A$. Now, increasing the resonance frequency of cavity $C$, slow enough in time to be adiabatic, the photons from cavity $A$ transferred to cavity $B$. In the experiment, the maximum transfer efficiency was 90\%. 
\begin{figure*}
\centering
\includegraphics[width=11cm,height=6.7cm]{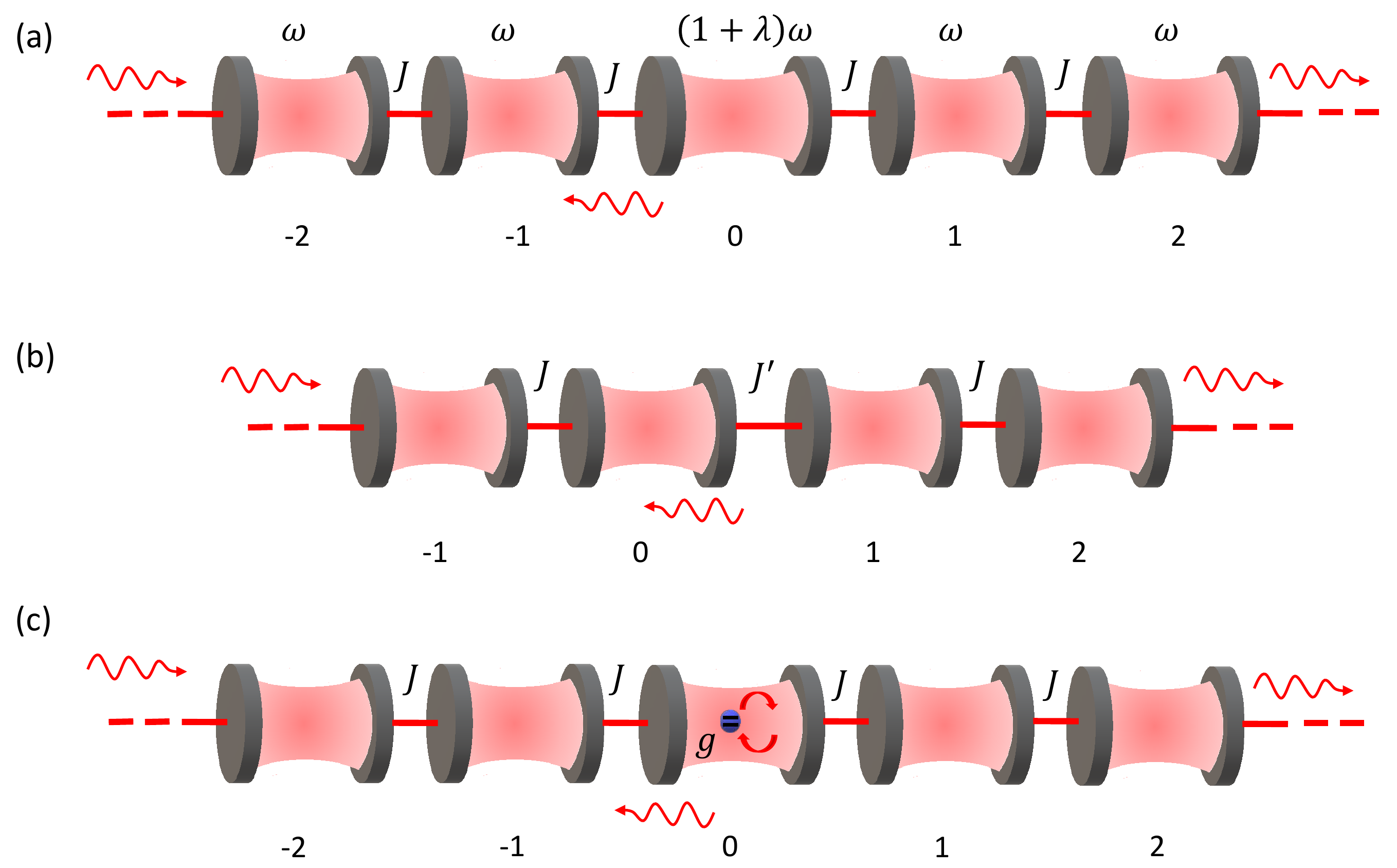}
\caption{Schematic of various array configurations for controlling photon transfer. (a) Middle cavity (0th cavity) of the homogeneously coupled cavity array is detuned from rest of the cavities by an amount of $\lambda\omega$. (b) The coupling strength between 0th cavity and 1st cavity is $J'= J+\lambda J$ while other cavities are coupled with their nearest cavities with the strength $J$. All the resonance frequencies of the cavities are equal. (c) The middle cavity of the homogeneously coupled cavity array is interacting with a two-level atom with the strength $g$. All the resonance frequencies of the cavities are equal. }
\label{Controllingtransfer}
\end{figure*}

In an infinite array, tuning the resonance frequency of one of the cavities in the array, the reflection (transmission) of a photon towards the left (right) in the array can be controlled. The Hamiltonian of the array shown in Fig. \ref{Controllingtransfer}(a) can be written as
\begin{align}
\hat H= \omega \sum_{j} \hat a_j^\dagger  \hat a_j +\lambda \omega \hat a_0^\dagger  \hat a_0+J \sum_{j} (\hat a_j^\dagger  \hat a_{j+1}+\hat a_j  \hat a_{j+1}^\dagger).
\end{align}
The frequency of the middle cavity is detuned by an amount $\lambda\omega$. Using the discrete scattering method \cite{Zhou2008PRL}, one can derive the single-photon reflection coefficient to be \cite{Liao2010PRA}
\begin{align}
R=\frac{(\lambda\omega)^2}{4J^2\sin^2 k+(\lambda \omega)^2},
\end{align}
where $k$ is the wave vector of the incident photon. As can be seen from the above equation, if there is no detuning $(\lambda=0)$, then the reflection coefficient $R$ becomes zero and the photon is completely transmitted from one end to the other end. However, the reflection probability is controllable by tuning the parameter $\lambda$, essentially the resonance frequency of the tuned cavity (0th cavity). In the same setup, if two cavities of the array are tuned, a supercavity is prepared that traps the photons in between the tuned cavities \cite{Liao2010PRA}.

Similarly, instead of tuning the resonance frequency, tuning any one of the coupling strengths in a homogeneously coupled cavity array, one can control the transfer of a photon \cite{Liao2009PRA}. Let the Hamiltonian for the array shown in Fig. \ref{Controllingtransfer}(b) be
\begin{align}
\hat H=& \omega \sum_{j} \hat a_j^\dagger  \hat a_j+ J \sum_{j} (\hat a_j^\dagger  \hat a_{j+1}+\hat a_j  \hat a_{j+1}^\dagger)\nonumber\\
&+\lambda J (\hat a_0^\dagger  \hat a_{1}+\hat a_0  \hat a_{1}^\dagger).
\end{align}
As can be seen, the coupling strength between 0th cavity and 1st cavity is $J'=J+\lambda J$, which can be tuned by changing the value of $\lambda$. Thus, $\lambda=(J'-J)/J$. Using the scattering method, the reflection coefficient calculated to be \cite{Liao2009PRA}
\begin{align*}
R=\frac{\lambda^2 (\lambda+2)^2}{\lambda^2 (\lambda+2)^2+4(\lambda+1)^2\sin^2 k}.
\end{align*}
Note that, $R$ is unity for $\lambda =-1$, indicating perfect reflection and zero transmission. In this case, the tuned coupling strength becomes zero and the cavity array becomes two separated arrays and therefore, no transfer is possible. For $\lambda=0$, $R$ becomes zero and complete transmission occurs $(T=1-R=1)$. Hence, the reflection coefficient $R$ can be tuned from 0 to 1 by tuning the coupling strength through $\lambda$.

It is also possible to control the transfer of photons by placing an atom in any one of the cavities in the array \cite{Shi2009PRB,Shi2011PRA,Qin2016PRA, Yan2014PRA,Shi2015PRA,Alexanian2010PRA,Zheng2010PRA,
Yan2012PRA,Shen2007PRL,Shen2007PRA,Longo2012OptExp,Stolyarov2020PRA}. The presence of an atom in the array either reflects or transmits the photon depending on the detuning between the atom and cavity \cite{Zhou2008PRL,Shen2005PRL,Cheng2012PRA}. Thus, the atom acts as a switch. Consider an array of cavities in which the middle cavity (0th cavity) contains a two-level atom, as can be seen in Fig. \ref{Controllingtransfer}(c). The Hamiltonian for this configuration is
\begin{align}
\hat H=& \omega \sum_{j} \hat a_j^\dagger  \hat a_j+J \sum_{j} (\hat a_j^\dagger  \hat a_{j+1}+\hat a_j  \hat a_{j+1}^\dagger)\nonumber\\
&+\omega_0\ket{e}\bra{e}+g(\hat a_0^\dagger \ket{g}\bra{e}+\hat a_0\ket{e}\bra{g}).
\end{align}
The single-photon reflection coefficient is \cite{Zhou2008PRL}
\begin{align}
R=\frac{g^4}{4J^2 \Delta^2\sin^2 k+g^4},
\end{align}  
where $\Delta=\omega-\omega_0-2J\cos k$ is the effective detuning and $g$ is the atom-cavity coupling strength.  For $\Delta=0$, photon is completely reflected and the two-level atom behaves as a perfect mirror. The reflection coefficient can be controlled by tuning $\Delta$. Therefore, the two-level system can be used as a quantum switch to control the coherent transport of photons. On the other hand, if the atom is strongly coupled to the middle cavity of the array in which the RWA picture breaks down, then the transport property becomes quite different than that of the weak coupling case \cite{Burillo2014PRL, Felicetti2014PRA,Wang2012PRA}. In the ultra-strong coupling regime, an incident photon deposits energy into the atom and escapes with a lower frequency \cite{Burillo2014PRL}. The efficiency of this nonlinear process is 50$\%$. 

If the atom simultaneously interacts with two nearest cavities resonantly, then the array acts like a Mach-Zehnder interferometer \cite{Zhou2012PRA}. As a result, the transmission and reflection spectra show interference pattern due to the lack of path information upon photon transfer. However, if the coupling between the atom and the two cavities is dispersive, then the reflection coefficient depends on the state of the atom \cite{Qin2016InJMPB}. For instance, if the atom is in its excited state, then the photon gets completely reflected, whereas the photon gets transmitted if the atom is in its ground state.

\begin{figure*}
\centering
\includegraphics[width=12cm,height=6.3cm]{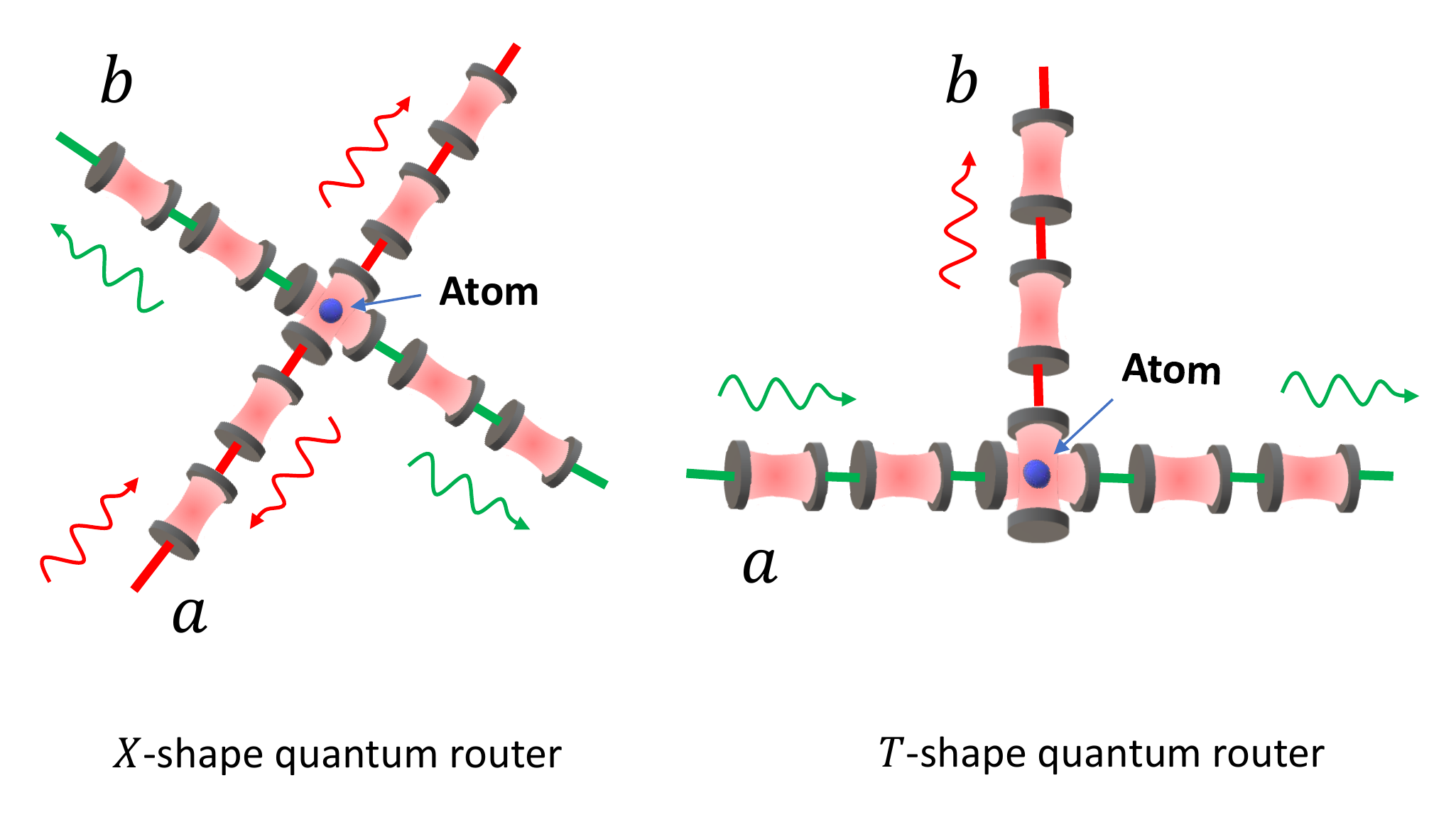}
\caption{Schematic of $X$-shape (two infinite arrays are coupled via an atom) and $T$-shape (an infinite and a semi-infinite arrays are coupled via an atom) quantum routers.   }
\label{QuantumRouter}
\end{figure*}

Extending the ideas of controlling the reflection and transmission of a photon by an atom in an array, quantum routers \cite{Aoki2009PRL,Lu2014PRA,Hoi2011PRL,
Yan2016EPJD,Ahumada2019PRA,Huang2018QIP,Huang2017JPhyB,Huang2019QIP,
Shi2018EPJD,Yan2016IJTP,Liu2019JLowTemp,Bao2019EPJD,Du2021EPJD,
Ren2022PRA} are proposed by coupling two different cavity arrays in $X$-shape \cite{Zhou2013PRL}, $T$-shape \cite{Lu2015OptExp,Shi2019OptComm,Liu2019OptExp}, $T$-bulge-shape \cite{Liu2016QIP,Zhang2021ChPhyB}, $\Pi$-shape \cite{Zhang2019IJTP}, etc. Routing of photons allows to connect several quantum nodes to build a quantum network. Schematic of $X$-shape and $T$-shape quantum routers are shown in Fig. \ref{QuantumRouter}. The $X$-shape quantum router, proposed in Ref. \cite{Zhou2013PRL}, consists of two infinite-dimensional cavity arrays and the middle cavities of both the array are simultaneously coupled to a cyclic three-level atom ($\Delta$-type) which is driven by a classical laser field. Consider a wave with energy $E$ is incident from one side of the array $a$. Using the scattering approach, one gets the transmittance to the array $b$ to be \cite{Zhou2013PRL}
\begin{align}
T_b=\left\vert\frac{2iJ_a \sin k_a G}{[2iJ_a \sin k_a-V_a][2iJ_b \sin k_b-V_b]-G^2}\right\vert^2,
\end{align}
where $J_a$ and $J_b$ are the coupling strengths between the cavities in the arrays $a$ and $b$ respectively. Also, $V_a=E g_a^2/(E^2-\Omega^2), V_b=E g_b^2/(E^2-\Omega^2)$ and $G=\Omega g_a g_b/(E^2-\Omega^2)$, where $g_a$, $g_b$ are the coupling strengths between the atom and the corresponding cavity of the arrays $a$ and $b$ respectively. $\Omega$ is the driving strength of the laser field. As can be seen, in the absence of driving ($\Omega=0$), the transmittance $T_b$ becomes zero. This indicates that the photon from the array $a$ cannot flow to the array $b$. However, in the presence of driving, the non-zero value of $T_b$ indicates the photon routing from one array to the other array. Hence, the classical field redirects the photon into another channel. But one can see that the maximum rate of transfer does not exceed 0.5, which may limits the complete transfer of quantum information from a sender to a receiver. In order to enhance the transfer rate, $T$-shape \cite{Lu2015OptExp,Shi2019OptComm,Liu2019OptExp} quantum router is proposed. In $T$-shape quantum router \cite{Lu2015OptExp}, one of the arrays is infinite-dimensional and other is semi-infinite (refer Fig. \ref{QuantumRouter}). The end cavity of the semi-infinite array and the middle cavity of the infinite array couple to a two-level atom. If a photon is incident to the array $a$, then the transmittance to the array $b$ is \cite{Lu2015OptExp}
\begin{align}
T_b=\left\vert\frac{-2g_a g_b \sin k}{2J \sin k (E-\omega_0)+g_b^2 \sin 2k+i(g_a^2+2g_b^2\sin^2 k)}\right\vert^2,
\end{align}
which gives the maximum rate of transfer is 0.5. Here $k=k_a=k_b$, and $\omega_0$ is the atomic transition frequency. However, when the photon is incident to the semi-infinite array $b$, then the transmittance to the array $a$ is \cite{Lu2015OptExp}
\begin{align}
T_a=2\left\vert\frac{-2g_a g_b \sin k}{2J \sin k (E-\omega_0)+g_b^2 \sin 2k+i(g_a^2+2g_b^2\sin^2 k)}\right\vert^2,
\end{align}
In this case, $T_b$ becomes unity indicating complete routing of photon from semi-infinite array to the infinite array. Hence, unidirectional high transfer probability of the incident photons is achieved from the semi-infinite channel to the infinite channel. But, the opposite direction transfer remains to be less than 0.5. There are  other type quantum routers such as multi-$T$-shape \cite{Huang2018QIP,Huang2018JPhyB}, $\Pi$-shape \cite{Zhang2019IJTP}, $T$-bulge-shape \cite{Liu2016QIP,Zhang2021ChPhyB}, six-port quantum router \cite{Tian2017OptComm}, asymmetrically coupled-cavities four-port quantum router \cite{Li2020CommThPhys} etc. are proposed for increasing the transfer rate to both sides.

The transport properties of photons in a finite number of cavities containing Kerr nonlinearity are also investigated for understanding the dynamics of strongly correlated photons \cite{Biella2015PRA,Heuvk2020PRA,Manosh2018JNOPM}. The transfer of photons in such arrays can be controlled by manipulating the coupling strengths \cite{Ferretti2010PRA} or phase of the coupling strengths \cite{Shen2015PRA}. By suitably choosing the Kerr strengths, a single photon can be transferred between any two cavities in the array without populating the intermediate cavities \cite{Meher2017ScRep}.


\section{Quantum information processing in cavities}
\begin{figure*}
\includegraphics[height=6cm,width=12cm]{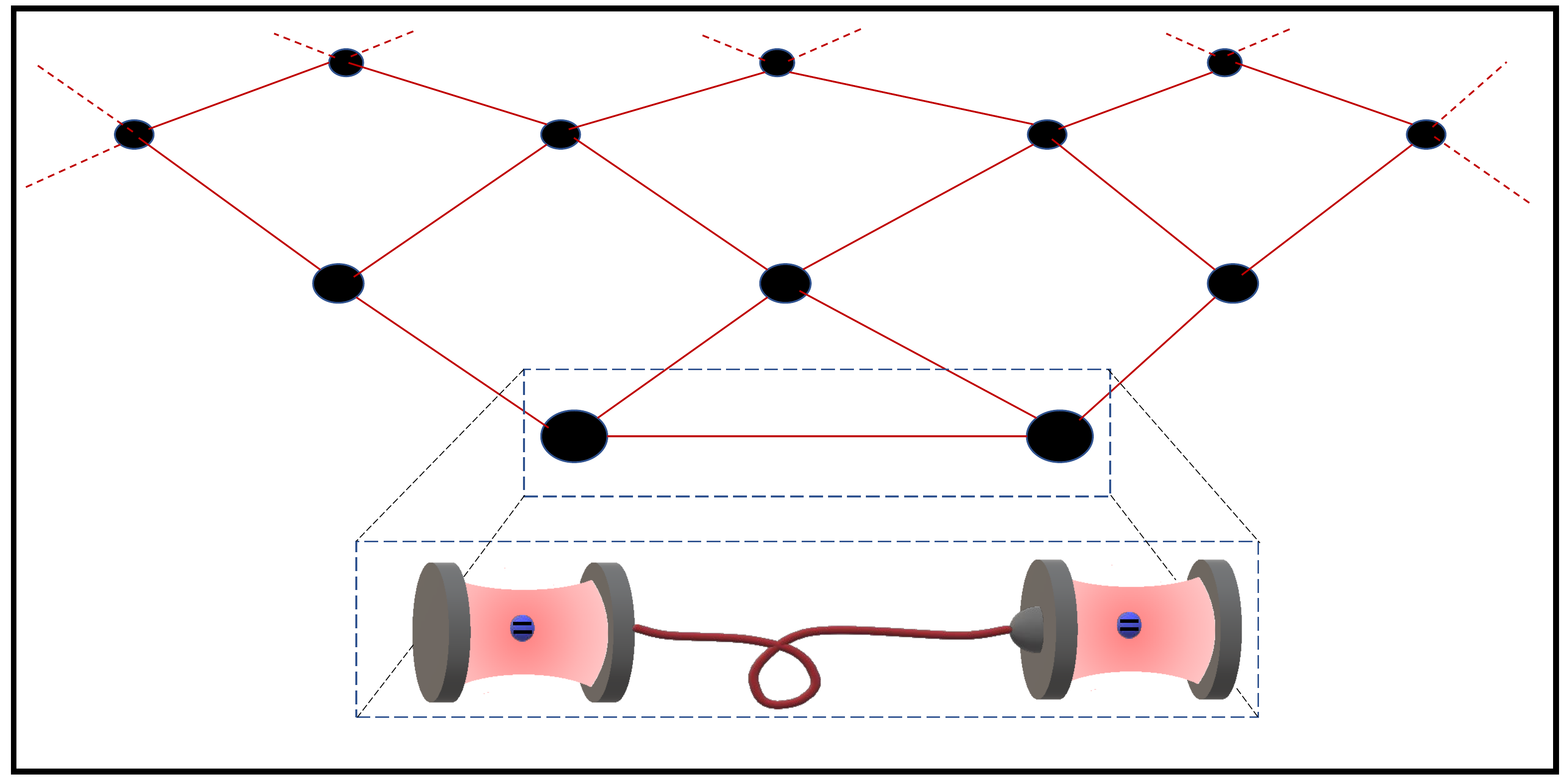}
\caption{ Schematic of a cavity-based quantum network. Quantum nodes (cavity with atom) are spatially separated and connected by optical fibers. }
\label{CavityNetwork}
\end{figure*}

Processing of quantum information can be realized in various physical systems such as spin chains, cavity array, array of trapped ions, etc. \cite{Cirac1998PhysScr,DiVincenzo2000FP,Bose2003PRL,Enk2004QInP,
Zhang2005PRA,Petrosyan2009PRA,Hijlkema2007NatPhys,Neil,Northup2014NatPhy,
Saffman2010RevModPhys,Blais2020NatPhys,Yang2007CommThPhy}. Among them, coupled cavity array provides an excellent setup for distributed quantum information processing and quantum communication, as it can be arranged in various locations and linked via optical fibers to form a quantum network  (see Fig. \ref{CavityNetwork}).  A quantum network must be capable
of sending and receiving information, creating entanglement, and performing quantum gate operations \cite{Cirac1997PRL, Marcos1999JOSAB, Lin2009APL, Reiserer2015RevModPhys}.    This section reviews a few important theoretical proposals for quantum information tasks and their experimental demonstrations in cavities.

\subsection{Quantum state transfer}
Quantum state transfer is essential for transferring information encoded in a quantum state from sender to a receiver \textit{via} a quantum channel. The sender prepares a quantum state $\ket{\psi}$ which encodes the information to be communicated to the receiver. The state $\ket{\psi}$ can be realized as the state of a qubit or a qudit ($d$- dimensional state). The receiver, after receiving the state $\ket{\psi}$, reads out the state for the information.

Let the sender wants to send a state $\ket{\psi}$ from the first cavity to the receiver which is located at the end cavity of the array. Then we can write the initial state of the array to be 
\begin{align}
\ket{\Psi}=\ket{\psi}_S\ket{0}\ket{0}....\ket{0}\ket{0}_R.
\end{align}
Here \textquoteleft$S$\textquoteright{} stands for sender and \textquoteleft$R$\textquoteright{} stands for receiver. If the state of the array after time \textquoteleft$T$\textquoteright{} becomes
\begin{align}
\ket{\Psi(T)}=\ket{0}_S\ket{0}\ket{0}....\ket{0}\ket{\psi}_R ,
\end{align}
then the perfect quantum state transfer is realized.

Suppose, the sender wants to send a qubit state $\ket{\psi}=\cos\frac{\theta}{2}\ket{0}+e^{i\phi}\sin\frac{\theta}{2}\ket{1}$ and the receiver wants to retrieve this state, or a state as close to it as possible, from the end cavity. Here, $\ket{0}$ represents the vacuum state and $\ket{1}$ is the single-photon state. Then, we write the initial state to be 
\begin{align}
\ket{\Psi(0)}=\cos\frac{\theta}{2}\ket{\text{vac}}+e^{i\phi}\sin\frac{\theta}{2}\ket{1}\rangle,
\end{align}
where $\ket{\text{vac}}=\ket{0}\ket{0}...\ket{0}$ and 
$\ket{1}\rangle=\ket{1}\ket{0}....\ket{0}$. After evolving this state under the Hamiltonian given in Eqn. (\ref{Ch2arrays}), the state at a later time \textquoteleft$t$\textquoteright{} is
\begin{align}
&\ket{\Psi(t)}=\cos\frac{\theta}{2}\ket{\text{vac}}+e^{i\phi}\sin\frac{\theta}{2}\sum_{j=1}^N C_{j1} \ket{j}\rangle,
\end{align}
where $C_{j1}=\langle\bra{j} e^{-i \hat Ht}\ket{1}\rangle$ is the single-photon transfer amplitude between 1st cavity and $j$th cavity. Here, $\ket{j}\rangle$ is the state representing single photon in $j$th cavity and the other cavities in their respective vacuum. The state of the receiver cavity (end cavity) is calculated by tracing over the states of all the other cavities, which yields \cite{Bose2003PRL}
\begin{align}
\rho_N(t)=P(t)\ket{\psi'}\bra{\psi'}+(1-P(t))\ket{0}\bra{0},
\end{align}
where $\ket{\psi'}=\frac{1}{\sqrt{P(t)}}\left(\cos\frac{\theta}{2}\ket{0}+e^{i\phi}\sin\frac{\theta}{2} C^*_{N1}\ket{1}\right)$ and $P(t)=\cos^2\frac{\theta}{2}+\sin^2\frac{\theta}{2} \vert C_{N1}\vert^2$. The term $C_{N1}=\langle\bra{N} e^{-i \hat Ht}\ket{1}\rangle$ is the transfer amplitude of single photon from the first cavity to $N$th cavity. The fidelity of quantum state transfer is calculated by averaging over all the pure initial states $\ket{\psi}$ in the Bloch sphere \cite{Bose2003PRL}. Therefore, 
\begin{align}\label{Averagefidelity}
{F}(t)=&\frac{1}{4\pi}\int\bra{\psi}\rho_N(t)\ket{\psi}d\Omega \nonumber\\
=&\frac{1}{2}+\frac{\vert C_{N1}(t)\vert \cos\gamma}{3}+\frac{\vert C_{N1}(t)\vert^2}{6},
\end{align}
where $\gamma=\text{arg}\{C_{N1}(t)\}$ which can be a multiple of $2\pi$ if the resonance frequencies and coupling strengths are properly chosen. This result can be generalized by replacing $C_{N1}$ by $C_{rs}=\langle\bra{r} e^{-i \hat Ht}\ket{s}\rangle$ in the context of transferring a quantum state from $s$th cavity to $r$th cavity in the array \cite{Bose2003PRL}.
 
Ideally, the transfer of a quantum state in an array of two cavities is possible with unit fidelity as $\vert C_{N1}\vert $ (here $N=2$) becomes unity \cite{Meher2017ScRep,Meher2020Arxiv}. However, dissipation and decoherence of the quantum states, which are not avoidable in experiments, limit the transfer fidelity. Recently, Axline \textit{et al.} \cite{Axline2018NatPhys} experimentally demonstrated a quantum state transfer protocol, based on a protocol proposed in Ref.  \cite{Cirac1997PRL}, between two superconducting microwave cavities connected by a transmission line. The sender cavity emits the state as a wavepacket with a specified temporal profile into the transmission line, and the receiver cavity absorbs this wavepacket to get the state. They could transfer a single-photon state and a coherent state of small amplitude between two cavities with a fidelity of $0.87 \pm 0.04$ \cite{Axline2018NatPhys}. 

\begin{figure}
\centering
\includegraphics[width=9cm,height=5cm]{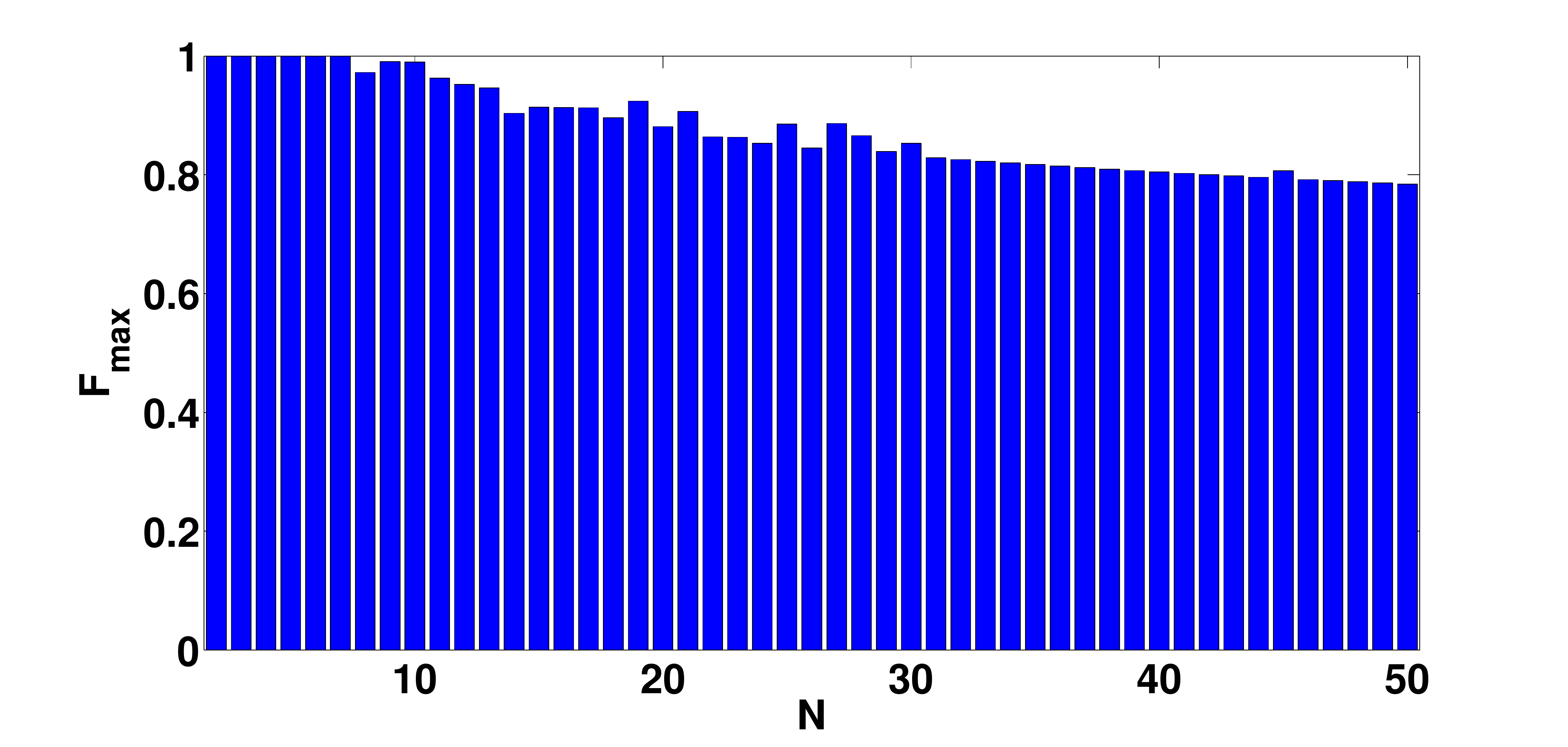}
\caption{Maximum value achieved by $F$ during time evolution within the interval $[0,2000/J]$ as a function of number of cavities in the array $N$ from 2 to 50.}
\label{FmaxVsNHomogeneous}
\end{figure}

If we include more cavities in the array, the transfer fidelity depends on the choices of coupling strengths. For an array of $N$ homogeneously coupled cavities described by the Hamiltonian given in Eqn. (\ref{HomogeneousArray}),  the single-photon transfer amplitude from Eqn. (\ref{averagephotonhomogeneous}) is
\begin{align}
C_{N1}(t)=\frac{2}{N+1}
\sum_{k=1}^N e^{-i\Omega_k t}\sin\left(\frac{N\pi k}{N+1}\right)\sin\left(\frac{\pi k}{N+1}\right),
\end{align}
where $\Omega_k=\omega+2J\cos\left(\frac{\pi k}{N+1}\right)$ is the normal mode frequency. Using Eqn. (\ref{Averagefidelity}), the average fidelity is calculated as a function of time and the peak value of fidelity (denoted by $F_{max}$) during a time interval from 0 to 20000/$J$ is shown in Fig. \ref{FmaxVsNHomogeneous}.  For $N=$2 and 3, the peak value is unity, and very close to unity for $N=4$ to 7, and it decreases as the number of cavities in the array increases. This indicates that the perfect transfer of a single-photon state does not occur if the array contains more than three cavities in a homogeneously coupled cavity array \cite{Christandl2004PRL,Godsil2012PRL,Meher2017ScRep}. 

Although the homogeneously coupled cavity array does not provide perfect state transfer, the array permits pretty good state transfer, that is, the fidelity of transfer becomes arbitrarily close to 1 if the number of cavities in the array is $N=p-1, 2p-1$, where $p$ is a prime, or $N=2^m-1$ \cite{Godsil2012PRL}. A pretty good state transfer can be achieved by tuning the coupling strengths and/or resonance frequencies \cite{Meher2020IJTP,YangLiu2015NJP,YaoPRL2011}. For instance, if the two end cavities are loosely coupled to the array \cite{YangLiu2015NJP,YaoPRL2011}, modelled by setting $J_1=J_{N-1}<<J_l$ for $l \neq 1,N-1$ in Eqn. (\ref{Ch2arrays}), the transfer fidelity of a quantum state becomes close to unity. But the time of perfect state transfer becomes arbitrarily large \cite{YangLiu2015NJP}.
Pretty good state transfer is also possible in Glauber-Fock cavity array \cite{Meher2020IJTP}, whose coupling strengths satisfy a square-root law $J_l=\sqrt{l}J$ \cite{Perez2010OptLetter,Longhi,Rai,Keil}. The interaction term of the Hamiltonian given in Eqn. (\ref{Ch2arrays}) becomes $\hat H_{int}=\sum_{l=1}^{N-1} 
\sqrt{l}J(\hat{a}_l^\dagger \hat{a}_{l+1}+\hat{a}_l \hat{a}_{l+1}^\dagger)$. 
By properly choosing the resonance frequencies, the single-photon transfer amplitude can be calculated to be \cite{Meher2020IJTP} 
\begin{align}
C_{N1}\approx -i\sin\theta t,
\end{align}
where $\theta$ depends on the resonance frequencies and coupling strengths, and becomes small for large $N$ \cite{Meher2020IJTP}. 
The time of transfer becomes large if the number of cavities in the array is large. The average fidelity becomes independent of the number of cavities in a staggered coupled cavity array \cite{Almeida2016PRA}, whose alternate coupling strengths are $J_{1,3,5,..}=(1+\eta)J$ and $J_{2,4,6,..}=(1-\eta)J$ in Eqn. (\ref{Ch2arrays}). For $J_{1,3,5,..}\ll J_{2,4,6,..}$, the single-photon transfer amplitude becomes \cite{Almeida2016PRA}
\begin{align}
\vert C_{N1}\vert \approx \left\vert \sin (\frac{\delta\omega}{2} t)\right\vert,
\end{align}
where $\delta\omega$ is the energy gap between the localized bound states. Although, this configuration does not provide perfect state transfer, the fidelity of state transfer is close to unity.

It is to be noted that all the aforementioned choices of coupling strengths do not allow perfect transfer of a photon as the transfer probability amplitude $\vert C_{N1}\vert $  given in Eqn. (\ref{Averagefidelity}) does not become unity. However, the coupling strengths given in Eqn. (\ref{sitedependentcoupling}), that is, $J_l=\sqrt{l(N-l)}J$, give  \cite{Perez2013PRA,Meher2017ScRep}
\begin{align}
C_{N1}=\sin^{N-1}Jt,
\end{align}
which becomes unity when time $t=\pi/2J$ for an arbitrary array length \cite{Meher2017ScRep,YangLiu2015NJP}. Hence, this particular array provides a perfect state transfer between two end cavities. It is also noted that the perfect state transfer is possible between two symmetrically located cavities \cite{Perez2013PRA,Meher2017ScRep}.  
Such an inhomogeneous coupling has been achieved in arrays containing a few cavities and a single-photon qubit state is transferred from $n$th to $(N-n+1)$th cavities \cite{Matthieu2012OptLett,Perez2013PRA,
Perez2013PRAFeb}. In Ref. \cite{Matthieu2012OptLett}, the reported fidelity is 0.65 with nine cavities. Higher fidelity of 0.84 has been reported in arrays with nineteen cavities \cite{Perez2013PRA,Perez2013PRAFeb}. In such experiments, the information can be encoded in a superposition of vacuum and single-photon states.  By implementing similar choice of coupling strengths in an array of 11 waveguides, polarization states of photons are transferred between two symmetrically located waveguides with a fidelity of 0.976$\pm$0.006 \cite{Chapman2016NatComm}. This form of coupling strengths also allow for perfect transfer of entangled states \cite{Perez2013PRAFeb,Chapman2016NatComm}.  

Quantum information can be encoded in a superposition of atomic states of the form $\alpha\ket{e}+\beta\ket{g}$ and can be preserved for a long time by placing the atom in a cavity. The atom-cavity system forms a quantum node, and linking many quantum nodes using optical fibers forms a quantum network (see fig. \ref{CavityNetwork}). Hence, transferring an atomic qubit state between two nodes is essential for large-scale quantum communication.  The atom whose state is to be transferred is called as sender and the atom that receives the state is referred to as receiver. Initially, the sender atom will be prepared in a qubit state $\alpha\ket{e}+\beta\ket{g}$ and will be allowed to interact with a cavity which is in vacuum $\ket{0}$. The evolved state of the atom-cavity system after a time $t$ becomes $\ket{g}(\alpha\ket{0}+\beta\ket{1})$, that is, the state of the atom gets mapped to the cavity field state. Now, if we send another atom which is in the ground state through the cavity, and properly choose the time of interaction, then the state of the atom after a time $t$ becomes $\alpha\ket{e}+\beta\ket{g}$.  Maitre \textit{et al.} experimentally demonstrated this state transfer protocol with a fidelity of 0.74 \cite{Maitre1997PRL}. Essentially, this experiment shows that a cavity can serve as a mediator to transfer quantum information between two atoms.  However, this scheme is not suitable for distance communication that requires the nodes to be spatially separated.  Cirac \textit{et al.} proposed a scheme to transfer a quantum state between two separated nodes through an optical fiber \cite{Cirac1997PRL}. This scheme uses special laser pulses that excite the sender atom so that its state is mapped into a time-symmetric photon wave packet that will enter to another cavity and to be absorbed by the receiver atom with unit probability. Implementing this scheme in an experiment, Kurpiers \textit{et al.} \cite{KurpiersNature2018} transferred a quantum state between two superconducting transmon qubits coupled to two coplanar microwave resonators separated by a distance of 0.9 m. The fidelity of state transfer was (80.02$\pm$0.07)\%. In another experiment, Ritter \textit{et al.} \cite{Ritter2012Nature} demonstrated the quantum state transfer between two distant nodes separated by a distance of 21 m and   connected by an optical fiber of 60 m length. Here the state of the sender atom initially gets mapped to the polarization state of a photon $\alpha \ket{L}+\beta \ket{R}$. The states $\ket{L}$ and $\ket{R}$ represent the left and right circular polarization states of the photon. Then the photon is transferred to another cavity through the fiber. The receiver atom receives the information from the photon by absorbing it. The fidelity of state transfer was 0.84. Importantly, these schemes are scalable to arbitrary network configurations of multiple atom-cavity systems; hence, large-scale quantum communication is possible. Several schemes for quantum state transfer between two distant atoms are theoretical proposals by several researchers and details can be found in the Refs. \cite{Enk1997JModOpt,Li2019OptExp,
Vogell2017QST,Nohama2007JModOpt, Bose2007JModOpt, Almeida2016PRA,LiJian2008CommThPhys,LiJian2010CommThPhy,
Zheng2016QInp,Rahman2011JAP,Yabu2013QInP,Badshah2020JOSAB, Manosh2018JNOPM}. In some of these protocols, information is encoded in the combined states of the atom and photon, so called polaritonic state. In such cases, the interaction between the atoms and cavities is considered to be resonant. However, there are a few schemes that use the non-resonant coupling between the atoms and cavities to transfer a quantum state between the atoms \cite{Zhou2009PRA,Ogden2008PRA,Kurizki1996PRA}. This state transfer is effected without populating the intermediate cavities. These schemes have the advantage that cavity decay can be avoided. However, a consequence of this non-resonant interaction is that the time of transfer is large and the transfer protocol may not be suitable for distant communication because of the unavoidable effect of atomic dissipation and decoherence. 
\subsection{Entanglement generation}
Quantum entanglement in the context of pure states is defined as a state of two or more quantum systems that cannot be expressed as a tensor product of the states of individual systems. When two systems are entangled,
each of them can reveal information about the other. Entanglement is essential for teleportation \cite{Bennet1993PRL}, dense coding \cite{Bennett}, cryptographic key distribution \cite{Bennett1992JourCrypt,Yin2020Nature}, etc. Quantum teleportation uses the non-local features of the entangled states to transmit a quantum state of a particle to another particle, whereas dense coding uses the entanglement correlation property to encode more information. Much effort has gone into generating entanglement in atom-cavity \cite{Raimond2001RevModPhys,Hagley1997PRL,Weber2009PRL,Solano2003PRL,
Aguilar2006PhysScr,Mohamed2015EPJD}, atom-atom \cite{Cirac1994PRA,Gerry1996PRA,Raimond2001RevModPhys,
Freyberger1995PRA,Yu2008EPJD,Zheng2000PRL,Osnaghi2001PRL,
Ritter2012Nature,Hagley1997PRL,Li2017OptLett,Mandilara2007PRA}, cavity-cavity \cite{Alexanian2011PRA,Tan2011PRA, Ikram2000OptComm,Chen2011OptComm,
Napoli2002JourModOpt,Liew2013NJP,Mohamed2018JOSAB,Mohamed2018QInP,
Mohamed2017PhysScr}, etc.

There are entangled states that can be prepared in two cavities by  sharing a single photon between them. These states are of the form
\begin{align}\label{Bellpsi}
\ket{\psi_{\pm}}=\frac{1}{\sqrt{2}}(\ket{1,0} \pm\ket{0,1}),
\end{align}
which are the Bell states \cite{Bell1964PPF} having maximum entanglement. To prepare these states in two cavities, one needs to couple a cavity being in a single-photon state $\ket{1}$ to another cavity which is in vacuum \cite{Meher2017ScRep}. By making the coupling strength to be complex, such that the interaction Hamiltonian is $\hat H_{int}=J(e^{i\eta} \hat a_1^\dagger \hat a_2+ e^{-i\eta} \hat a_1 \hat a_2^\dagger)$ \cite{Meher2017ScRep}, the evolved state of the coupled cavities becomes $\cos Jt\ket{1,0}-ie^{-i\eta}\sin Jt\ket{0,1}$. At time $t=\pi/4J$, the evolved state is the entangled state in Eqn. (\ref{Bellpsi}) for $\eta=\mp\pi/2$. In general, one can generate entangled states of the form $\cos\theta\ket{1,0}+e^{i\phi}\sin \theta\ket{0,1}$. There are also other methods to prepare the Bell states given in Eqn. (\ref{Bellpsi}). For instance, recently, Axline \textit{et al.} \cite{Axline2018NatPhys} experimentally generated the state $\ket{\psi_+}$ between two distant cavities by sending a single-photon wavepacket through them. The wavepacket releases half of its stored energy in the first cavity, as a result, the emitted wavepacket and the first cavity get entangled. Then the subsequent absorption of the emitted wavepacket by the other cavity generates the desired entangled state between the cavities. The fidelity of generating the state $\ket{\psi_{+}}$ was 0.77 \cite{Axline2018NatPhys}. In another scheme, instead of sending a single-photon wavepacket, a two-level atom being in the excited state is sent through the cavities to prepare the above entangled states \cite{Rauschenbeutel2001PRA,Davidovich1994PRA}. By detecting the atom in its ground state after interaction with the cavities, the desired entangled state is generated in the cavities.  But this scheme is probabilistic because it fails if the atom is detected in its excited state \cite{Browne2003PRL}. The states $\ket{\psi_{\pm}}$ can also be prepared in two cavities by sending a three-level atom through the cavities \cite{Ikram2002PRA}. 

If there are total $N$ photons in two cavities, a possible entangled state is of the form
\begin{align}\label{photondistribution}
\ket{\psi}= \sum_{k=0}^N c_k \ket{k,N-k}.
\end{align} 
Essentially, $N$ photons are distributed between the cavities, and the probability of detecting $k$ photons in the first cavity and the remaining photons in the second cavity is $\vert c_k\vert^2$. Theoretical proposals for generating such entangled states are given in Refs.  \cite{Ikram2000OptComm,Wildfeuer2003PRA}. According to these schemes,  one needs to send $N$ number of excited two-level atoms through two cavities to generate this state. The interaction between the atom and cavity is the resonant Jaynes-Cummings interaction. By
controlling the atom-cavity interaction times, one can generate the entangled state given in Eqn. (\ref{photondistribution}). In this scheme, one needs to suitably choose $2N$ parameters (two parameters for each atom) such that all the atoms are found to be in their ground states after
crossing the cavities. If any one of the
$N$ atoms is detected to be in excited state then the scheme fails. Hence, this scheme is good for generating entangled state with a small number of photons such that the detection of atoms in excited state can be avoided. A special case of the state given in Eqn. (\ref{photondistribution}) is 
\begin{align}\label{Ch2EntState}
\ket{\Psi}=\cos\theta\ket{m,n}+e^{i\phi}\sin\theta\ket{p,q},
\end{align}
where $m+n=p+q=N$. This state can be prepared by including Kerr nonlinearity in two coupled cavities \cite{Meher2017ScRep} by choosing the resonance frequencies of the cavities and Kerr nonlinearity strengths suitably. For $n=q=N$, $m=p=0$, $\theta=\pi/4$ and $\phi=0$, the above state becomes the NOON state
\begin{align}
\ket{NOON}=\frac{1}{\sqrt{2}}(\ket{N,0}+\ket{0,N}).
\end{align}
These states are special since the Heisenberg limit is achievable in phase-sensitivity \cite{Bollinger1996PRA}. Wang \textit{et al.} have experimentally demonstrated the generation of NOON states in two cavities \cite{Wang2011PRL}. In their experiment, two superconducting phase qubits and three microwave cavities are used. One of the cavities is used for coupling both the qubits and other two cavities are used for state storage. According to their experimental scheme, to generate the NOON state with $N$=1, which is the Bell state given in Eqn. (\ref{Bellpsi}), one needs to swap the qubit Bell state $\ket{eg}+\ket{ge}$ with the storage cavities, that is,
$\ket{eg}+\ket{ge}\ket{00}\rightarrow \ket{gg}(\ket{10}+\ket{01}).$
To generate higher-order NOON states, we need another higher excited atomic level $\ket{f}$ and prepare the entangled state $\ket{fg}+\ket{gf}$. The transition from $\ket{f}\rightarrow \ket{e}$ generates the atom-cavity entangled state $\ket{eg10}+\ket{ge01}$. The qubits are excited to their $\ket{f}$ states again, and  the transition $\ket{f}\rightarrow \ket{e}$ prepares the state $\ket{eg20}+\ket{ge02}$. Following the same procedure, one can generate the state upto $(N-1)$ photons, that is, $\ket{eg(N-1)0}+\ket{ge0(N-1)}$. At the final step, $\ket{e}\rightarrow \ket{g}$ transition of both the qubits generates the state $\ket{ggN0}+\ket{gg0N}$, which is the NOON state in the cavity modes while both the qubits are in their ground states. NOON states up to three photons were generated with fidelities 0.76$\pm$0.02 $(N=1)$, 0.50$\pm$0.02 $(N=2)$ and 0.33$\pm$0.02 $(N=3)$. The fidelity degrades rapidly as the number of photons increases because of the dissipation and experimental difficulties. There are also theoretical proposals for generating NOON states of higher values of $N$ with high fidelities in cavities \cite{Meher2017ScRep, Nikoghosyan2012PRL, Strauch2010PRL,Kamide2017PRA,Merkel2010NJP}.   
 
In all the entangled states in the previous discussions, the number of photons distributed between two cavities is fixed. However, there are entangled states in which the number of photons to be distributed between two cavities is not fixed. For example, 
\begin{align}\label{BellPhi}
\ket{\phi_{\pm}}=\frac{1}{\sqrt{2}}(\ket{0,0}\pm\ket{1,1}),
\end{align}
which are also Bell states \cite{Bell1964PPF}. We will get either no photons or one photon in each cavity on measurement and hence, this state cannot be generated by distributing a fixed number of photons between two cavities. Ikram \textit{et al.} \cite{Ikram2002PRA} theoretically suggested that the state $\ket{\phi_\pm}$ can be generated by sending a three-level atom (V-type atom with two excited states $\ket{a},\ket{b}$ and a ground state $\ket{c}$) through two cavities. The first cavity is resonant with $\ket{a}\leftrightarrow \ket{c}$ transition and the second cavity is resonant with $\ket{b}\leftrightarrow \ket{c}$ transition. The initial state of the system is $\ket{a,0,0}$: the atom is in the state $\ket{a}$ and the two cavities are in vacuum. After resonant interaction between the atom and first cavity, the evolved state at a particular time will be $\frac{1}{\sqrt{2}}(\ket{a,0,0}+\ket{c,1,0})$. Now, before entering to the second cavity, the atom is driven by a laser pulse for the transition $\ket{c}\rightarrow \ket{b}$. The state of the system becomes $\frac{1}{\sqrt{2}}(\ket{a,0,0}+\ket{b,1,0})$. Then the atom enters to the second cavity and the transition $\ket{b}\leftrightarrow \ket{c}$ produces the state $\frac{1}{\sqrt{2}}(\ket{a,0,0}+\ket{c,1,1})$. As can be seen, this is an entangled state between the atom and two cavities. However, for a time larger than the lifetime of the atom, the state collapses to the entangled state given in Eqn. (\ref{BellPhi}). The state $\ket{\phi_-}$ can be generated if the relative difference of interaction times of atoms with the two
cavities is $\pi$ \cite{Ikram2002PRA}. 

A more general entangled state in $\{\ket{0},\ket{1}\}$ basis is
\begin{align}
\ket{\psi}=c_{00}\ket{0,0}+c_{01}\ket{0,1}+c_{10}\ket{1,0}+c_{11} \ket{1,1}).
\end{align}
All the Bell states can be realized by suitably choosing the superposition coefficients. This state can be entangled or not depending on the superposition coefficients \cite{Neil}. This class of states can be prepared in two cavities using a single two-level atom interacting with two auxiliary classical fields \cite{Tiegang2004JMOpt}. After a conditional measurement on the atom, the cavity fields collapse to the desired entangled state. Though the fidelity for preparing this state is unity, the scheme is probabilistic because of the conditional measurement.  

Some of the higher dimensional bi-partite entangled states are also relevant for performing quantum information tasks. Among them, an important class of states is the entangled coherent states which is important for teleporting higher dimensional states \cite{vanEnk2003PRL}. The form of entangled coherent states is \cite{Sanders1992PRA,Davidovich1993PRL}
\begin{align}
\ket{\psi(\alpha,\beta)}=\frac{1}{\sqrt{N}}(\ket{\alpha}\ket{\beta}\pm \ket{-\alpha}\ket{-\beta}),
\end{align}
where $\alpha$ and $\beta$ are two complex amplitudes and $N$ is the normalization constant. These states can be prepared in two cavities by simultaneously interacting with a driven atom \cite{Solano2003PRL}.  In general, by sending sequence of atoms through two cavities containing two equal amplitude coherent states \cite{Zou2005EPJD}, one can generate multi-dimensional entangled coherent states
\begin{align}
\ket{\psi}=\frac{1}{\sqrt{N}}\sum_{j=0}^{N-1} C_j \ket{\alpha e^{-ij2\pi/N}}\ket{\alpha e^{ij2\pi/N}}.
\end{align}
These states are more entangled than entangled coherent states and can be used for quantum teleportation \cite{vanEnk2003PRL}. These superposition states can approximate pair coherent states \cite{Agarwal1986PRL} and superposition of pair coherent states \cite{Gerry1995PRA}.

Multi-partite entangled states are relevant for processing more amount of quantum information. Two important classes of multi-partite entangled states are W states and Greenberger-Horne-Zeilinger (GHZ) states. The form of W state is
\begin{align}
\ket{\psi_W}=\frac{1}{\sqrt{3}}(\ket{1,0,0}+\ket{0,1,0}+\ket{0,0,1}).
\end{align}
This state is robust against qubit loss. As can be seen, if one qubit is lost, the state of the remaining qubits is still entangled. On the other hand, a multi-qubit W state exhibits stronger nonclassical behavior than a GHZ state \cite{SenDe2003PRA} and hence, it is an ideal resource for
communication based on multinodal networks \cite{Zhang2006PRA},
teleportation and dense coding \cite{Agrawal2006PRA} and optimal universal quantum cloning \cite{Murao1999PRA}. As can be seen, the W state can be prepared by sharing a photon in three cavities. It is theoretically suggested that the W state can be prepared by sending an excited two-level atom through three identical cavities \cite{Bilal2012JMOpt,Yang2004IJQInf,Miry2015QInP}. Detecting the atom in its ground state, the desired entangled states is produced in the cavity and the scheme fails if the atom is detected in the excited state. The $N$-party W state can also be prepared in a similar manner where one needs to send an excited atom through $N$ cavities \cite{Zainub2020LaserPhys}. Lee \textit{et al.} proposed a scheme to prepare the W state in cavities by swapping the atomic W state \cite{Lee2008PRA}. 

Another multi-partite entangled state is the GHZ state, which is a maximally entangled state, of the form
\begin{align}
\ket{\psi_{GHZ}}=\frac{1}{\sqrt{2}}(\ket{0,0,0}+\ket{1,1,1}).
\end{align}
Entanglement in the state will be destroyed upon qubit loss. On measurement, either we do not get photons in any of the cavities or we get one photon in each cavity. Using Jaynes-Cummings and anti-Jaynes-Cummings interactions between atoms and cavities, Miry \textit{et al.} have shown the possibility of generating the GHZ state in three and four cavities \cite{Miry2015QInP}.  However, in a simpler way, Farook \textit{et al.} \cite{Bilal2012JMOpt} theoretically showed that this state can be prepared by sending two two-level atoms through three cavities among which the first cavity should be initiated in the state $\frac{1}{\sqrt{2}}(\ket{0}+\ket{3})$. This scheme can be extended to prepare $N$ partite GHZ state by considering the first cavity to be in the state $\frac{1}{\sqrt{2}}(\ket{0}+\ket{N})$ \cite{Zainub2020LaserPhys}. 


\subsection{Realization of quantum gates}
Quantum gates are the basic ingredients for building a quantum computer \cite{Benioff1980JStatPhys,Feynman1982IJTP,DiVincenzo4, DiVincenzo3, Ladd, Barz, Cai}. Deutsch and Josza showed that a quantum computer could solve some classes of computational problems much faster than a classical computer \cite{Deut}. To solve a desired problem in a quantum computer, one needs to build a quantum circuit by suitably arranging various quantum gates \cite{Shor, Fredkin, DiVincenzo2, Bare}. However, if one identifies a set of universal quantum gates, a minimal set of gates can be used for constructing the same circuit. A set of gates are said to be universal if any gates can be constructed using the set \cite{Fredkin, DiVincenzo2,Bare, Sleator, Lloy, David, Buzek2,Shi, Neil,DiVincenzo4,DiVincenzo3}. For instance, few single-qubit gates with controlled-NOT (CNOT) gate, \textit{i.e,} $\{$CNOT, Hadamard, phase and $\pi/8 (T)\}$, collectively forms a universal set.


\begin{table*}
\caption{Reported quantum gate fidelities in experiments}
\begin{center}
 \begin{tabular}{|| c | c | c ||} 
 \hline
References & Gates & Fidelity  \\ [0.5ex] 
 \hline\hline
Kandala \textit{et al.} \cite{Kandala2021PRL}& CNOT gate & 99.77 $\%$ \\
\hline
Rosenblum \textit{et al.} \cite{Rosenblum2018NatComm}& CNOT gate & (90 $\pm$ 2) $\%$ \\
  \hline
  Premaratne \textit{et al.} \cite{Premaratne2019PRA}& CNOT gate & 84 $\%$ \\
  \hline
Reiserer \textit{et al.} \cite{Reiserer2014Nature}& CNOT gate & 80.7 $\%$ \\
  \hline
Welte \textit{et al.} \cite{Welte2018PRX}& CNOT gate &  74.1 $\%$ \\
 \hline
  Hacker \textit{et al.} \cite{Hacker2016Nature}& CNOT gate & (76.9 $\pm$ 1.5)$\%$ \\ 
 \hline
 Hacker \textit{et al.} \cite{Hacker2016Nature}& Controlled phase-flip gate & (76.2 $\pm$ 3.6)$\%$ \\ 
 \hline
Sung \textit{et al.}\cite{Sung2021PhysRevX}  & iSWAP gate & (99.87 $\pm$ 0.23$)\%$ \\
 \hline
 Sung \textit{et al.}\cite{Sung2021PhysRevX}  & CZ gate & (99.76 $\pm$ 0.07)$\%$ \\
 \hline
 Li \textit{et al.}\cite{Li2019npjQI}  & CZ gate & (99.54 $\pm$ 0.08)$\%$ \\
 \hline
 Li \textit{et al.}\cite{Li2019npjQI}  & CCZ gate & 93.3$\%$ \\
 \hline
\end{tabular}
\label{GateFidelity}
\end{center}
\end{table*}

As the qubits are represented by an arbitrary superposition of $\ket{0}$ and $\ket{1}$, the single-qubit gate operation is represented by a $2\times 2$ matrix.  The unitary operator for a general single-qubit gate is of the form \cite{Fredkin, DiVincenzo2,Bare, Sleator, Lloy, David, Buzek2,Shi, Neil}
\begin{align}
U(\theta,\phi,\lambda)=\left(\begin{array}{cc}
\cos\left(\frac{\theta}{2}\right) & -e^{i\lambda}\sin\left(\frac{\theta}{2}\right) \\ e^{i\phi}\sin\left(\frac{\theta}{2}\right) & e^{i\phi+i\lambda}\cos\left(\frac{\theta}{2}\right)
\end{array}\right),
\end{align} 
called the U-gate. One can get various important single-qubit gates like the Hadamard gate
\begin{align}
H=U(\pi/2,0,\pi)=\frac{1}{\sqrt{2}}\left(\begin{array}{cc}
1 & 1 \\ 1 & -1
\end{array}\right),
\end{align} 
X-gate
\begin{align}
X=U(\pi,0,\pi)=\left(\begin{array}{cc}
0 & 1 \\ 1 & 0
\end{array}\right),
\end{align}
phase-flip gate
\begin{align}
Z=U(0,\pi/2,\pi/2)=\left(\begin{array}{cc}
1 & 0 \\ 0 & -1
\end{array}\right),
\end{align}
phase gate
\begin{align}
S=U(0,\pi/4,\pi/4)=\left(\begin{array}{cc}
1 & 0\\ 0 & i
\end{array}\right),
\end{align}
and $\pi/8$ gate 
\begin{align}
T=U(0,\pi/8,\pi/8)=\left(\begin{array}{cc}
1 & 0\\ 0 & e^{i\pi/4}
\end{array}\right)
\end{align}
by suitably choosing the parameters $\theta,\phi$ and $\lambda$.

Single-qubit gates are easy to implement in a two-level atom by applying suitable Rabi pulses \cite{Sleator,Giovannetti2000PRA,
Goto2004PRA,He2010CommThPhys}. But it is not so easy in cavity field modes because they have infinite energy levels. However, there are a few schemes for realizing single-qubit gates in cavities by sending atoms through the cavity and applying additional electromagnetic pulses for suitable atomic transitions \cite{Giovannetti2000PRA}. By suitably choosing the interaction time, single-qubit gates such as the X-gate, Hadamard gate can be realized.  The X-gate and phase gate can also be realized by controlling frequency, phase and amplitude of the drive to the cavity in an atom-cavity system \cite{Blais2007PhysRevA}. If coherent states are considered as qubits, Kerr nonlinearity in the cavity is required to realize quantum gates \cite{Goto2016PRA}.

A collection of single-qubit gates cannot form a universal set to build any desired quantum circuit. Two-qubit gates are essential for building a universal set \cite{Fredkin, DiVincenzo2,Bare, Sleator, Lloy, David, Buzek2,Shi, Neil,Saffman2010RevModPhys}. Therefore, any operation on qubits can be broken up into a concatenation of two-qubit operations accompanied by single-qubit operations \cite{DiVincenzo2}. Some important two-qubit gates are the CNOT gate, quantum phase gate, SWAP gate, etc., to mention a few. In a two-qubit operation, one of the qubits is considered as the control bit and the other is the target bit. If a two-qubit gate operates on a two-qubit state, the state of the target qubit changes depending on the state of the control bit. These gates are represented by $4\times 4$ matrices. 

CNOT gate coherently operates on a two-qubit state of the form  $c_{00}\ket{00}+c_{01}\ket{01}+c_{10}\ket{10}+c_{11}\ket{11}$. The result of CNOT gate operation is $\ket{a}\ket{b}\rightarrow \ket{a}\ket{a \bigoplus b}$, with $a,b \in \{0,1\}$ and $\bigoplus$ represents the addition modulo 2. Here, the state $\ket{a}$ is considered as the control bit and $\ket{b}$ is the target bit. In CNOT operation, if the state of the control bit is $\ket{0}$ then the state of the target bit remains the same. Flipping occurs between $\ket{0}$ and $\ket{1}$ of the target bit if the state of the control bit is $\ket{1}$. Hence, the unitary operator for the CNOT gate in the basis $\{\ket{00}, \ket{01}, \ket{10},\ket{11}\}$ is
\begin{align}
U_{CNOT}=\left(\begin{array}{cccc}
1 & 0 & 0 & 0\\
0 & 1 & 0 & 0\\
0 & 0 & 0 & 1\\
0 & 0 & 1 & 0
\end{array} \right).
\end{align}
This gate has several properties and applications, as discussed in   Ref. \cite{Barenco1995PRL}. For example, the CNOT gate transforms some superpositions into entangled states and, thus, it acts as a measurement gate. This gate can also be used to implement Bell measurement and swapping operation. The CNOT gate is not a universal
gate by itself. Together with a few single-qubit gates, it forms a universal set \cite{Barenco1995PRL}. CNOT gate has been realized in many physical systems experimentally. In the atom-cavity system, there are a few notable experimental realizations of CNOT gate that we discuss here \cite{Welte2018PRX,Premaratne2019PRA,
Rosenblum2018NatComm,Kandala2021PRL,Hacker2016Nature}. An advantage of  realizing quantum gates in an atom-cavity system is the freedom in choosing the control bit and target bit. For instance, Rosenblum \textit{et al.} \cite{Rosenblum2018NatComm} demonstrated the CNOT gate operation by encoding qubits in the high-dimensional space of the cavity field. The controlled qubits are
\begin{align}
\ket{0}_C=\ket{0}_C, \ket{1}_C=\ket{2}_C,
\end{align} 
which are even-parity Fock states and the target qubits are
\begin{align}
\ket{0/1}_T=\frac{1}{\sqrt{2}}\left(\frac{\ket{0}_T+\ket{4}_T}{\sqrt{2}}\pm \ket{2}_T\right),
\end{align}
the Schrodinger kitten states \cite{Rosenblum2018NatComm, Michael2016PRX}. These qubits are robust against photon loss and they allow for error correction in the target cavity. A basic requirement for realizing this gate in their setup is the nonlinear interaction between the cavities, which is achieved by making the cavities to interact sequentially with a transmon qubit. In the experiment, the CNOT gate was applied on the state $(\ket{0}_C+\ket{1}_C)\ket{0}_T$. The output  was the Bell state $\ket{0}_C\ket{0}_T+\ket{1}_C\ket{1}_T$ generated with a fidelity about (90 $\pm$ 2)\%. The gate time in their experiment was 190 ns. Similarly, Hacker \textit{et al.} \cite{Hacker2016Nature} also considered the photonic states as the qubits and encoded the information in polarization states of photons, and demonstrated the CNOT gate operation in a one-sided optical cavity containing a Rubidium atom. The gate fidelity was $(76.9 \pm 1.5)\%$. The scheme requires one of the photons to be circularly polarized (control bit) and the other linearly polarized (target bit). On the other hand, Welte \textit{et al.} \cite{Welte2018PRX} and  Premaratne \textit{et al.} \cite{Premaratne2019PRA} realized CNOT gate between two atoms, by encoding the qubit states in the atomic states, with a fidelity of 74.1\% and 84\% respectively. The gate time for the former experiment was 2 $\mu$s and the latter was 907 ns. It is also advantageous to choose both atomic and photonic states as the qubits. For instance, Reiserer \textit{et al.} \cite{Reiserer2014Nature} used the spin states of the atom as the control bit and polarization states of the photon as the target bit, and demonstrated the CNOT gate experimentally with a fidelity of $80.7$\%. This scheme is deterministic and robust, and applies to almost any matter qubit. The reported values of fidelities of CNOT gate in various experiments are listed in Table. \ref{GateFidelity}. There are many theoretical proposals to realize the CNOT gate operations in cavities \cite{Sleator,Giovannetti2000PRA, Domokos,Biswas2004PRA, Meher2019JPhysB,Wang2016ScRep} involving nonlinear mechanical interaction \cite{Meher2019JPhysB}, adiabatic passage \cite{Sangouard2005EPJD,Zhe2016JModOpt}, atom-cavity dressed states \cite{Domokos}, ion-cavity interaction \cite{Feng2002JOptB}, non-local transformations \cite{Paternostro2003JModOpt}, non-resonant interaction of atom-cavity \cite{Yang2006PhysicaA,Tang2009CommThPhys}, etc.

Some of the quantum algorithms require changing the phase of the target bit depending on the state of the control bit. A gate that performs the conditional phase shift is a quantum phase gate. The operation is represented by \cite{Turchette1995PRL, Rauschenbeutel1999PRL}
\begin{align}
\ket {a,b} \rightarrow exp(-i\phi \delta_{a,1}\delta_{b,1})\ket{a,b},
\end{align}
where $\delta_{a,1}$ and $\delta_{b,1}$ are Kronecker delta symbols and $a,b \in \{0,1\}$. Hence, the phase of
the target bit changes by phase $\phi$ if the states of both control and target bits are $\ket{1}$. The unitary matrix for the quantum phase gate in the basis $\{\ket{00},\ket{01},\ket{10},\ket{11}\}$ is
\begin{align}
U_{QPG}=\left( \begin{array}{cccc}
1 & 0 & 0 & 0\\
0 & 1 & 0 & 0\\
0 & 0 & 1 & 0\\
0 & 0 & 0 & e^{-i\phi} 
\end{array} \right).
\end{align}
The quantum phase gate with single-qubit gates forms a universal set \cite{Rauschenbeutel1999PRL, Neil}. Turchette \textit{et al.}  have experimentally demonstrated the quantum phase gate by exploiting the nonlinear character of atom-cavity interaction \cite{Turchette1995PRL}. The phase shift between circular polarization states of atoms depends on the intensity of a pump beam via a Kerr-type nonlinear interaction. The conditional phase shift was nearly $16^0$ per intracavity photon and the range of phase shift was $-30^0$ to $30^0$. The range of the phase shift is further increased between $54^0$ to $273^0$ in the experiment by Rauschenbeutel \textit{et al.} \cite{Rauschenbeutel1999PRL}, wherein the qubits were represented by atomic states and cavity field states, and the atom-cavity dispersive interaction performs the quantum phase gate operation.
In addition, there are proposals that use the idea of Stark-shifted Raman transitions \cite{Biswas2004PRA}, photon-number Stark-shift \cite{Zou2007PRA}, adiabatic passage \cite{Goto2004PRA}, resonant interaction \cite{Zheng2005PRA}, dispersive interaction  \cite{Dong2015PhysLettA}, Kerr interaction \cite{Heuck2020PRL, Hiroo2008JPhyDAppPhy, Meher2019JPhysB}, multi-level interaction \cite{Xiao2006PRA}, etc. for shifting the phase, thereby enabling the construction of quantum phase gate.   
 
The quantum phase gate with $\phi=\pi$ is called controlled phase-flip gate or controlled-Z gate. The corresponding unitary representation is
\begin{align}
U_{CZ}=\left( \begin{array}{cccc}
1 & 0 & 0 & 0\\
0 & 1 & 0 & 0\\
0 & 0 & 1 & 0\\
0 & 0 & 0 & -1 
\end{array} \right).
\end{align}
This gate is important because a CNOT gate can be constructed from
one controlled phase-flip gate and two Hadamard gates \cite{Neil}. Hacker \textit{et al.} \cite{Hacker2016Nature} experimentally demonstrated this gate in a one-sided high-finesse optical cavity containing a Rubidium atom. The atom acts as an ancilla qubit and the polarization states $\ket{L}$ (left circularly polarized) and $\ket{R}$ (right circularly polarized) of both the photons are used as control bit and target bit. This experimental realization was based on the proposal by Duan and Kimble \cite{Duan2004PRL}. In the experiment,  the gate transformed $\ket{RR}\rightarrow \ket{RR}$, $\ket{LR}\rightarrow -\ket{LR}$, $\ket{RL}\rightarrow \ket{RL}$, $\ket{LL}\rightarrow \ket{LL}$, that is, introduced a phase $\pi$ if the control photonic bit is in the state $\ket{R}$ and target photonic bit is in $\ket{L}$. The average fidelity of the gate operation was (76.2 $\pm$ 3.6)\%. There are a few theoretical proposals to realize controlled phase-flip gates in atom-cavity system \cite{Yin2007PRA,Yang2009PRA} and ion-cavity system \cite{Zou2002PRA,Semiao2002PRA}.

There are other few important quantum gates such as the SWAP gate \cite{Sung2021PhysRevX,Biswas2004PRA,Serafini2006PRL,
Yin2007PRA,Lin2008PRA,
LiJian2008CommThPhys,Song2010OptComm,Jiang2008ChinPhysB,
Xiao2009PhysScr,Deng2011ChinPhysB,Yan2018QIP,Zhang2020QIP}, $\sqrt{\text{SWAP}}$ gate \cite{Koshino2010PRA,Liu2007ChinPhysLett}, entangling gates \cite{Serafini2006PRL,Yin2007PRA,LiJian2008CommThPhys},  geometric phase gate \cite{Zheng2004PRA2,Chen2006PRA,Feng2007PRA}, which can be prepared in atom-cavity system, are important for building quantum circuits. However, to scale up the quantum computation and process more information, multi-qubit gates are better choices. Using the quantum nonlinear interaction between atom and cavity, Toffoli gate \cite{Giovannetti2000PRA,Chen2006PRA2, Tang2009ChinPhysB, Shao2009ChinPhysB,Chen2014JOSAB}, Fredkin gate \cite{Shao2009ChinPhysB2,Shao2014JOSAB,Song2015QIP}, multi-qubit quantum phase gates \cite{Zou2007PRA,Xiao2007PRA2,Chang2008PRA,Yang2008PhysLettA,
Shao2009OptComm,Tang2009ChinPhysLetters,Lu2010ChinPhysLett,Lu2010ChinPhysB,
Lin2009PRA,Fan2008ChinPhysLett},  etc. are proposed in cavity QED setup.
\subsection{Quantum teleportation}
Bennet \textit{et al.} showed that an arbitrary unknown one-qubit state could be transferred from one party to another distant party using a maximally entangled state as a channel. This process of information transfer is named as quantum teleportation. The teleportation protocol is as follows. Consider three particles A, B and C, out of which the sender has two particles A and B, and the receiver has C. Sender prepares particle A, whose state is to be teleported, in an unknown quantum state $\ket{\psi}$. The other two particles, namely, B and C are maximally entangled.  Sender performs a joint measurement on particles A and B, and communicates the result to the receiver through classical communication. After receiving this measurement result, the receiver applies an appropriate rotation to particle C. Following this rotation, the state of particle C becomes $\ket{\psi}$.

\begin{figure}
\centering
\includegraphics[width=9cm,height=7cm]{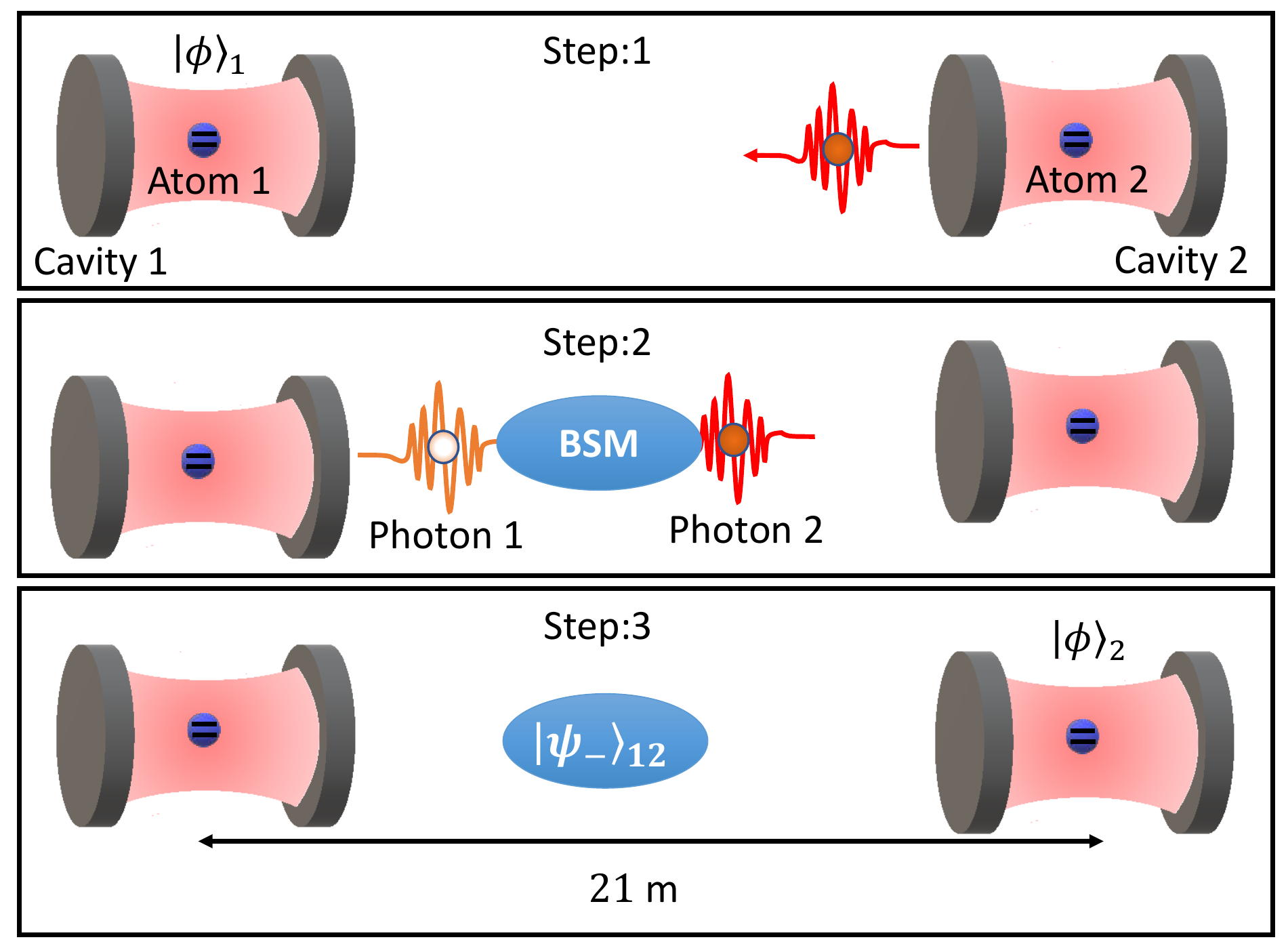}
\caption{A schematic of experimental setup for teleportation protocol by Nolleke \textit{et al.} \cite{Nolleke2013PRL}. Atom 1 and Atom 2 are trapped in two distant cavities separated by 21 m. A joint Bell state measurement (BSM) is performed on photons emitted from two nodes (atom-cavity). Teleportation is complete upon detection of $\ket{\psi_-}_{12}$ state. }
\label{TeleporatationNolleke}
\end{figure}
Teleportation experiments have been performed between two photonic qubits \cite{Bouwmeester1997Nature,Boschi1998PRL,
Kim2001PRL,Pan2003Nature,Lombardi2002PRL,Takeda2013Nature},
photonic and atomic qubits \cite{Davidovich1994PRA,Sherson2006Nature,Chen2008NatPhys, Bussieres2014NatPhotonics}, and two atomic
qubits \cite{Riebe2004Nature,Barrett2004Nature,Krauter2013NatPhys}. In cavity QED setup, Nolleke \textit{et al.} \cite{Nolleke2013PRL} have demonstrated the quantum teleportation experiment between two atoms trapped in two widely separated cavities. A schematic of their experimental setup is shown in Fig. \ref{TeleporatationNolleke}. Sender prepares Atom 1 in an unknown state $\ket{\phi}_1=\alpha \ket{e}+\beta \ket{g}$ and mapped this state to Photon 1. The receiver prepares an entangled state $\ket{\Psi^-}_{22}=\frac{1}{\sqrt{2}}(\ket{g}\ket{R}-\ket{e}\ket{L})$ between Atom 2 and  Photon 2 (emitted from cavity 2). The combined state of Photon 1, Photon 2 and Atom 2 is \cite{Nolleke2013PRL} 
\begin{align}
\ket{\phi}_1\ket{\psi_-}_{22}&=\frac{1}{2}(\ket{\Phi^+}_{12}\hat\sigma_{x}\hat\sigma_{z}\ket{\phi}_2-\ket{\Phi^-}_{12}\hat\sigma_{z}\ket{\phi}_2 \nonumber\\
&+\ket{\Psi^+}_{12}\hat\sigma_{x}\ket{\phi}_2-\ket{\Psi^-}_{12}\ket{\phi}_2),
\end{align}
where $\ket{\Phi^+}_{12},\ket{\Phi^-}_{12},\ket{\Psi^+}_{12}$ and $\ket{\Psi^-}_{12}$ are the Bell states of Photon 1 and Photon 2 in polarization basis.
Bell state measurement (BSM) is performed on Photon 1 and Photon 2. The teleportation is completed after detecting the state $\ket{\Psi_-}_{12}$. In their experiment, the fidelity of teleportation was 88.0$\pm 1.5$\%. Otherwise, one needs to apply a suitable rotation operation on atom. Very recently, Langenfeld \textit{et al.} \cite{Langenfield2021PRL} also experimentally demonstrated a teleportation protocol between two atoms  (separated by 60 m) with a fidelity of (88.3$\pm 1.3$)\%. 

In the literature, there are theoretical proposals to teleport a state between two atoms using atom-field entanglement \cite{Bose1999PRL,Zheng2004PRA, Zheng2005ChinPhys,Zheng1997PhysLettA} or atom-atom entanglement as the teleportation channels \cite{Cirac1994PRA2,Zheng2000PRL,Zheng1999OptComm,
Zheng2001PRA,Ye2004PRA,
Xiao2005PhysA,Zhuo2008PhysLettA}. Also, there are few schemes that consider the teleportation of an atomic state to a cavity field state \cite{Davidovich1994PRA} or teleportation between two cavities \cite{Moussa1996PRA,Liu2006PhysA}. For instance, Davidovich \textit{et al.} \cite{Davidovich1994PRA} used the original idea by Bennet \textit{et al.} and proposed an experimentally feasible scheme to teleport an atomic qubit state to a cavity field state by using the Bell state of two cavities as the teleportation channel. Most of these schemes involve Bell state measurement as in the original scheme proposed by Bennet \textit{et al.} \cite{Bennet1993PRL}. However, there are proposals, for instance, by Zheng \cite{Zheng2004PRA} and Yang \cite{Zhen2006JPhysB}, that do not involve Bell state measurement. Zheng \cite{Zheng2004PRA} showed that by sending two atoms through a single cavity, it is possible to teleport a state of one atom to another atom, but with a fidelity of less than unity. This scheme uses the time-dependent resonant interaction between the atoms and cavity. In the scheme by Yang \cite{Zhen2006JPhysB}, the atomic state can be teleported to another atom via resonant interaction between two atoms with two cavities. This scheme allows unit fidelity, but the success probability of the protocol is 1/4. Other issues that affect the fidelity of teleportation in the atom-cavity system are dissipation and decoherence of the atomic and cavity field states. To minimize the effect of dissipation and decoherence \cite{Zheng2001PRA,Yu2004PRA, Zheng2006PRA2,Zheng2005ChinPhys,Zhang2008,Zheng2008PRA,
ZhengBio2006CommThPhy}, atom-cavity dispersive interaction \cite{Zheng2000PRL,Zheng1999OptComm,Ye2004PRA,Xiao2005PhysA} or   detection of cavity decay photons \cite{Bose1999PRL,Yu2004PRA,Zheng2006PRA2,Cho2004PRA} can be exploited. The schemes involving dispersive atom-cavity coupling, in which the cavity is only virtually excited, become insensitive to cavity decay and therefore, maximize the teleportation fidelity \cite{Ye2004PRA}. However, the scheme by Bose \textit{et al.} \cite{Bose1999PRL} is based on the detection of the leakage photons from the cavity. Here the decay mechanism plays a constructive role and increases the teleportation fidelity.  Cirac \textit{et al.} \cite{Cirac1994PRA2} also proposed a teleportation protocol that has the advantage of minimizing dissipation in the optical regime, that is, in teleportation procedure, only ground state of atoms are ever populated. As a result, the spontaneous decay of the atoms is absent.

\begin{figure*}
\centering
\includegraphics[width=12cm,height=2.2cm]{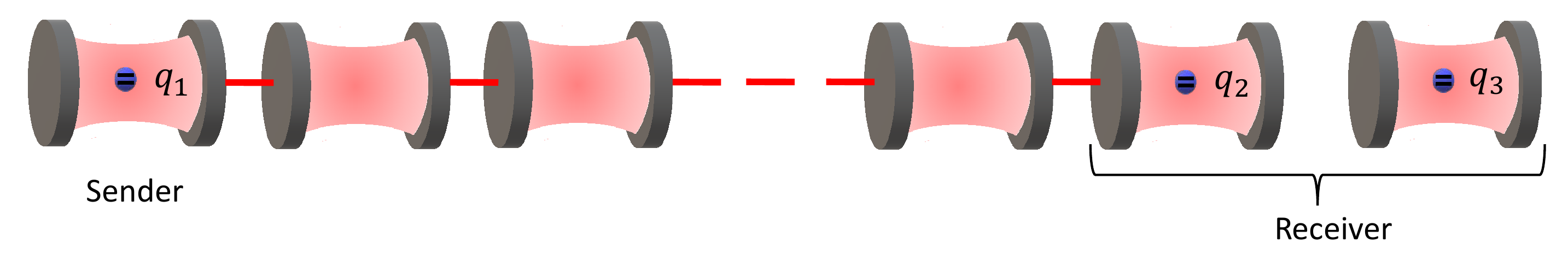}
\caption{Schematic of the proposed setup in Ref. \cite{Meher2020JPhysB} for quantum dense coding protocol. The setup consists of a cavity array whose end cavities contain one atom in each. The atom $q_1$ belongs to sender and the atom $q_2$ belongs to receiver. The receiver has an additional atom $q_3$ in a separate cavity from the array.}
\label{DenseCodingArray}
\end{figure*}

Information can also be encoded in higher dimensional states (qudits) and hence, the teleportation of a higher dimensional state is essential \cite{Moussa1998ModPhysLettB,Almeida2000PRA,Zheng1997PLA}. Moussa \textit{et al.} proposed a scheme to teleport a cavity field state $c_0 \ket{0}+c_1\ket{1}+c_2\ket{2}$ from one cavity to another cavity using two pairs of non-maximally entangled states \cite{Moussa1996PRA}. In general, superposition of $N$ Fock states $\sum_{k=0}^{N-1} c_k \ket{k}$ can be teleported using $N$ pairs of two-level atoms \cite{Moussa1996PRA}. However, as the scheme does not involve maximally entangled state, the maximum probability for completing the teleportation process is 0.5. Zubairy showed that the state $\sum_{k=0}^{N-1} c_k \ket{k}$ can be
teleported by using the following entangled field state \cite{Zubairy1998PRA} 
\begin{align}
\ket{\psi}=\frac{1}{\sqrt{N}}\sum_{k=0}^{N-1}\ket{N-1-k}\ket{k},
\end{align} 
as a quantum channel with unit fidelity. But this method is valid only if $N=2^n$ ($n$ is an integer).

To send more information, the sender needs to teleport multi-partite states. Proposals have been made to teleport multi-qubit  states such as the Dicke states \cite{Di2005PRA}, atomic entangled states \cite{Wang2007CommThPhys,Liu2006JOSAB, Yang2007ChinPhys}, entangled field states \cite{Ikram2000PRA, Pires2004PRA,Pires2005PRA,Cardoso2005PRA,dSouza2011OptComm,
Quarrat2008JPhyB}, etc.  These schemes require multi-qubit entangled states such as the GHZ state \cite{Almeida1998PLA,Ye2002ChinPhys} or W state \cite{Zhuo2004PhysA,Yuan2005ChinPhys} as quantum channels.
\subsection{Quantum dense coding}
Quantum dense coding protocol allows a sender to transmit two bits of classical information by sending only one qubit to the receiver  \cite{Bennett}. However, entanglement is essential for dense coding. The procedure is as follows. Initially, the sender and receiver share a maximally entangled state. Let the shared entangled state be $\ket{\psi_-}=\frac{1}{\sqrt{2}}(\ket{1,0}-\ket{0,1})$. Then, the sender applies one of the transformations from the set $\{ \hat I, \hat\sigma_x, i\hat\sigma_y, \hat\sigma_z\}$ on the entangled state to encode the corresponding two bits of classical information $(0,0), (0,1), (1,0)$ and $(1,1)$. After the transformation, the resultant states are
\begin{align}
\hat{I}\ket{\psi_-}=\ket{\psi_-},\\
\hat\sigma_x\ket{\psi_-}=\ket{\phi_-},\\
i\hat\sigma_y\ket{\psi_-}=\ket{\phi_+},\\
\hat\sigma_z\ket{\psi_-}=\ket{\psi_+}.
\end{align}
Then the sender sends its particle to the receiver, and the receiver performs a Bell state measurement. From the measurement outcome, the receiver can discriminate the operation sender has applied on his qubit and decode the classical information. 

Dense coding has been demonstrated experimentally in nuclear magnetic resonance \cite{Fang2000PRA}, optical systems \cite{Mattle,Li2002PRL}, etc. There are theoretical proposals to realize dense coding in other physical systems as well. However, there are a few experimentally feasible schemes have been proposed in the context of atom-cavity system, which either uses the Bell states \cite{Lin2003PhysLettA,Ye,Meher2020JPhysB} or multi-partitite entangled states \cite{Nie,Zheng2010ChinPhysB,Juan} such as the GHZ states \cite{Xue,Ye2005PhysLettA,Zou2011IntJourModPhyB,
Wang2007ChinPhysLett,Sun2014ChinPhysB,Li2013IJTP}, W-state \cite{Ye2005PhysLettA,Yu2008ChinPhysB,Zou2008CommThPhys,
Peng2008OptComm}, cluster states \cite{Jia2008CommThPhys,Yuan2013CommThPhys} etc. For example, Ye \textit{et al.} \cite{Ye} showed the protocol using two atoms dispersively interacting with a cavity. Initially, the sender and receiver shared the entangled state $\frac{1}{\sqrt{2}}(\ket{g,e}-i\ket{e,g})$. After applying one of the transformations from $\{ \hat I, \hat\sigma_x, i\hat\sigma_y, \hat\sigma_z\}$, the sender sends the atom to the receiver. The receiver lets both the atoms simultaneously interact with another cavity and also, the two atoms are driven by a classical field. By properly choosing the interaction time, the resultant state allows the receiver to discriminate the operation that the sender applied. The success probability of this scheme is unity. Also, the scheme is insensitive to the thermal field and cavity decay as it involves dispersive interaction between the atoms and cavity. There are a few proposals of quantum dense coding which also use the idea of dispersive interaction between the atom and cavity to suppress the decay of cavity \cite{Lin2003PhysLettA,Wu2008IJTP}. However, the detrimental effect like spontaneous emission of the atoms cannot be avoided in such schemes. Yu \textit{et al.} \cite{Yu2006JPhyB} have proposed a scheme based on the adiabatic passage and photonic interference, which makes the scheme more robust against certain types of noises such as atomic spontaneous emission, output coupling and inefficient photon detection.

In all the aforementioned schemes, the sender needs to send the atom to the receiver and hence, atoms are used as flying qubits. This may limit the transfer of information to a distant receiver. However, a scheme of quantum dense coding is proposed wherein the atoms can be used as stationary qubits and photons as the information carrier through a cavity array \cite{Meher2020JPhysB}. The schematic of the proposed setup is given in Fig. \ref{DenseCodingArray}. The sender has a single atom $(q_1)$ and the receiver has two atoms $(q_2$ and $q_3)$. The atom $q_1$ is entangled with the atom $q_3$ in the state $\frac{1}{\sqrt{2}}(\ket{e_1, e_3}+\ket{g_1,g_3})$, where $\ket{g_i}$ and $\ket{e_i}$ refer to the ground and excited states respectively of the atom $q_i$. The sender applies a unitary operation on $q_1$ depending on the choice of the classical bits to encode. After the operation, the sender allows the atom $q_1$ to interact resonantly with the first cavity of the array. Due to the interaction, the state of $q_1$ is transferred to $q_2$ through the cavity array, resulting in entanglement between the atoms $q_2$ and $q_3$. Now, the receiver discriminates the unitary operation applied by the sender upon a Bell measurement on $q_2$ and $q_3$. 
\section{Summary}
Quantum information processing requires physical systems wherein quantum features are created, sustained and manipulated to perform computational tasks \cite{Kimble2008Nature}.  A physical architecture to perform these tasks is composed of some basic units such as cavities \cite{Kimble2008Nature,Ritter2012Nature}, trapped ions \cite{Bruzewicz2019APR}, superconducting qubits \cite{Devoret2013Science}, etc. Among them, cavities have the potential to store multiple excitations of electromagnetic field and retain their coherence, and,  therefore, they facilitate the generation and manipulation of complex multiphoton states \cite{Vlastakis2013Science}.  These multiphoton states are resilient despite decoherence making them resources for quantum information processing \cite{Vlastakis2013Science}, quantum metrology \cite{Giovannetti2011NatPhot} and quantum computation \cite{Rosenblum2018NatComm}.  A  \textquotedblleft good\textquotedblright~ cavity or high-$Q$ cavity retains photons for a long time $\sim 1$ ms \cite{Raimond2001RevModPhys}. Recent technological progress in the fabrication of high-$Q$ cavities has rendered it possible to couple many cavities to build  photonic integrated circuits on chips \cite{Sato2011Nature}. Manipulation of trapped photons on the chip can be used to implement quantum protocols at a miniaturized scale, i.e., a huge number of components can be packed in a small spatial region. Such integrated photonic circuits have also wider applications such as creating slow light, buffering and switching of photons \cite{Sato2011Nature}, etc.

In the context of long distance communication, a large-scale network of cavities is required. An advantage of cavity-based architecture is that the coupling between distant cavities can be established via optical fibers,   a well-established technology developed for the conventional communication systems.  Coupling between the cavities allows to transfer quantum states between them with high fidelities and distribute entanglement across the entire network. This, in turn, enables distributed quantum information processing and communication. In addition, if a network contains an atom or nonlinear media in each of the cavities, controlled transfer of photons between a source cavity and target cavity is feasible due to this in-cavity nonlinearity. The in-cavity nonlinearity also allows for manipulation of cavity field states, generation of single photons, creation of entanglement among the fields in the cavities, etc. These physical processes provide scope for creating quantum gates, teleportation and dense coding in cavities. Therefore, a distributed quantum communication architecture is achievable in cavity networks which in all likelihood will be a suitable candidate for realizing quantum internet in near future \cite{Kimble2008Nature}.

Even though cavity networks are viable systems to carry out quantum information processing and computation, there are issues to be sorted out which limit their efficiency. At a practical level, making nearly identical cavities having suitable resonance frequencies, which are essential for strong coupling between the cavities, is a difficult task. However, tuning of resonance frequencies over a wide range is possible through various techniques \cite{Vignolini2010ApplPhysLett,Vignolini2010AppPhysLett2,
Fushman2007APL,DUndar2012APL,Hopman2006OptExp,Cai2013APL}.   Like any other physical system, cavities also suffer from dissipation which may lead to destruction of quantum coherence and thereby inhibit the transfer of states between the cavities. To sustain a  quantum state in the cavity for a long time, feedback mechanisms are necessary \cite{Vitali1998PRA}.  The loss of photons and accumulation of errors due to change of state of photons while passing through optical fibers are other factors that affect the efficiency of transferring  information in a network. It is, however, possible to circumvent this deficiency by incorporating quantum repeaters in conjunction with atom-based quantum memories in the network \cite{Sangourd2011RevModPhys,Ruihong2019JourPhysConf} which guarantees high fidelity transfer of quantum states between the sender and the receiver over long,  lossy channels.  There are also smart algorithms such as the quantum error correction \cite{Devitt2013RepProgPhys} to identify errors gathered in the quantum state of photons while passing through noisy channels and rectify the errors. These techniques are robust enough to perform highly reliable computations and long distance quantum communications.

Although, we have restricted our review to the relevance of cavities containing atoms or nonlinear medium for quantum information processing and computation,  thermal diodes \cite{Barzanjeh2018PRL, Meher2020JOSAB} and thermal transistors \cite{Xiong2018EPL} can be realized using cavities as conduits to transfer energy between two thermal baths.    A related physical system is an optomechanical device that has one or more cavities interacting with mechanical oscillators.  These are also suitable for studying the fundamentals of light-matter interaction and implementing quantum protocols  \cite{Aspelmeyer2014RevModPhys,Stannige2012PRL}. In optomechanical systems, photons and phonons are used as qubits to encode information.  Both these platforms, the  coupled cavities and the optomechanical systems,  are reliable for experimentally verifying ideas in quantum information processing, quantum thermodynamics \cite{Gelbwaser2013PRA,Gelbwaser2013EPL,Villa2018Quantum} and quantum chemistry \cite{Sidler2020JPhysChemLett}.  However, it is beyond the scope of this review to present optomechanics and quantum thermodynamics.

%

\end{document}